\documentclass[twocolumn]{aastex63}

\received{\today}
\revised{\today}
\accepted{\today}
\submitjournal{ApJ}

\shorttitle{Transport of charged particles through turbulent magnetic fields}
\shortauthors{Chen et al.}
\graphicspath{{./}{figures/}}

\begin{document}

\title{Transport of high-energy charged particles through spatially-intermittent turbulent magnetic fields}

\correspondingauthor{P.~Tzeferacos}
\email{petros.tzeferacos@flash.uchicago.edu}

\author{L.~E.~Chen}
\author{A.~F.~A.~Bott}
\affiliation{Department of Physics, University of Oxford, Parks Road, Oxford OX1 3PU, UK}

\author{P.~Tzeferacos}
\affiliation{Department of Physics, University of Oxford, Parks Road, Oxford OX1 3PU, UK}
\affiliation{Department of Astronomy and Astrophysics, University of Chicago, 5640 S. Ellis Ave, Chicago, IL 60637, USA}

\author{A.~Rigby}
 
\affiliation{Department of Physics, University of Oxford, Parks Road, Oxford OX1 3PU, UK}
 
\author{A.~Bell}
 
\affiliation{Department of Physics, University of Oxford, Parks Road, Oxford OX1 3PU, UK}
 
\author{R.~Bingham}
 
\affiliation{Rutherford Appleton Laboratory, Chilton, Didcot OX11 0QX, UK}
\affiliation{Department of Physics, University of Strathclyde, Glasgow G4 0NG, UK}

\author{C.~Graziani}
 \affiliation{Argonne National Laboratory,
 Mathematics and Computer Science Division, Argonne, IL, USA}

\author{J.~Katz}
 
\affiliation{Laboratory for Laser Energetics, University of Rochester, 250 E River Rd, Rochester, NY 14623, USA}

\author{M.~Koenig}
\affiliation{Laboratoire pour l'Utilisation de Lasers Intenses, UMR7605, CNRS CEA, France}
\affiliation{Graduate School of Engineering, Osaka University, Suita, Osaka 565-0871, Japan}

\author{C.~K.~Li}
 
\affiliation{Massachusetts Institute of Technology, 77 Massachusetts Ave, Cambridge, MA 02139, USA}

\author{R.~Petrasso}
 
\affiliation{Massachusetts Institute of Technology, 77 Massachusetts Ave, Cambridge, MA 02139, USA}

\author{H.-S.~Park}
 
\affiliation{Lawrence Livermore National Laboratory, 7000 East Ave, Livermore, CA 94550, USA}

\author{J.~S.~Ross}  
 
\affiliation{Lawrence Livermore National Laboratory, 7000 East Ave, Livermore, CA 94550, USA}

\author{D.~Ryu}
 
\affiliation{Department of Physics, School of Natural Sciences, UNIST, Ulsan 44919, Korea}
 
%\author{D.~Ryutov}
 
%\affiliation{Lawrence Livermore National Laboratory, 7000 East Ave, Livermore, CA 94550, USA}

\author{T.~G.~White}
\affiliation{Department of Physics, University of Oxford, Parks Road, Oxford OX1 3PU, UK}
\affiliation{Department of Physics, University of Nevada, Reno, Nevada 89557, USA}

\author{B.~Reville}
 
\affiliation{School of Mathematics and Physics, Queens University Belfast, Belfast BT7 1NN, UK}

\author{J.~Matthews}
 
\affiliation{Department of Physics, University of Oxford, Parks Road, Oxford OX1 3PU, UK}
 
\author{J.~Meinecke}
 
\affiliation{Department of Physics, University of Oxford, Parks Road, Oxford OX1 3PU, UK}
 
\author{F.~Miniati}
 
\affiliation{Department of Physics, University of Oxford, Parks Road, Oxford OX1 3PU, UK}

\author{E.~G.~Zweibel}
 
\affiliation{Departments of Astronomy and Physics, University of Wisconsin, 475 N. Charter Street, Madison, Wisconsin 53706, USA}
  
\author{S.~Sarkar}
 
\affiliation{Department of Physics, University of Oxford, Parks Road, Oxford OX1 3PU, UK}
%\affiliation{Niels Bohr Institute, Blegdamsvej 17, 2100 Copenhagen, Denmark}

\author{A.~A.~Schekochihin}

\affiliation{Department of Physics, University of Oxford, Parks Road, Oxford OX1 3PU, UK}

\author{D.~Q.~Lamb}
 
\affiliation{Department of Astronomy and Astrophysics, University of Chicago, 5640 S. Ellis Ave, Chicago, IL 60637, USA}

\author{D.~H.~Froula}
 
\affiliation{Laboratory for Laser Energetics, University of Rochester, 250 E River Rd, Rochester, NY 14623, USA}

\author{G.~Gregori}
 
\affiliation{Department of Physics, University of Oxford, Parks Road, Oxford OX1 3PU, UK}
\affiliation{Department of Astronomy and Astrophysics, University of Chicago, 5640 S. Ellis Ave, Chicago, IL 60637, USA}

\begin{abstract}
Identifying the sources of the highest energy cosmic rays requires
understanding how they are deflected by the stochastic, spatially intermittent
intergalactic magnetic field.  Here we report measurements of energetic
charged-particle propagation through a laser-produced magnetized plasma
with these properties. We characterize the diffusive transport of the particles
experimentally. The results show that the transport is diffusive and that,
for the regime of interest for the highest-energy cosmic rays, the diffusion
coefficient is unaffected by the spatial intermittency of the magnetic field.
\end{abstract}

\keywords{High energy astrophysics  --- Laboratory astrophysics --- Particle astrophysics --- Magnetic fields --- Intergalactic medium ---  Plasma astrophysics --- Ultra-high-energy cosmic radiation}

\section{Introduction} \label{sec:intro}

The interplay between charged particles and stochastic magnetic fields generated
by plasma turbulence is crucial to understanding how cosmic rays propagate through
space \citep{Strong2007,Zweibel2013,Schlickeiser2015}. A key parameter for determining
the underlying nature of charged-particle diffusion is the ratio of the particle
gyroradius $r_g$ to the correlation length $\ell_B$ of the magnetic turbulence.
For the vast majority of cosmic rays detected at the Earth, this ratio is small.
These are particles that are well confined by the Galactic magnetic field.
But for cosmic rays more energetic than about 10 EeV, the ratio is larger than unity.
These ultra-high-energy cosmic rays (UHECRs) are not confined to the Milky Way and
are presumed to be extragalactic in origin.  Identifying their sources requires
understanding how they are deflected by the intergalactic magnetic field, which
appears to be stochastic and spatially intermittent. Recent data from the Parker Solar Probe mission have also indicated the presence of non-Gaussian magnetic fields near the Sun \citep{Bandyopadhyay2019arXiv}.

To study the propagation of cosmic rays, a theoretical framework has been developed
based on direct numerical simulations of particle trajectories
(e.g., \citet{Sigl2003,Sigl2004}) and statistical techniques (see \citet{ShalchiBook2009}
for a review). In particular, it has been shown  \citep{Jokipii1966} that random,
small-amplitude fluctuations of the magnetic field superimposed on a mean background
field lead to diffusive particle propagation. As a result, standard (Markovian)
diffusion is widely used in modeling cosmic-ray transport (e.g., \citet{Kotera2008,
Globus2008, Globus2017, Globus2019}), although anomalous diffusion has been shown
to occur in special cases \citep{Jokipii1969, Reville2008, Lazarian2014},
including resonant scattering of charged particles in spatially intermittent
magnetic fields \citep{Shukurov}.

Past laboratory experiments have studied particle transport in diffuse plasmas with
strong mean magnetic fields \citep{Gustafson2012,Anderson2013,Furno2015,Bovet2015},
but the regime that is relevant to UHECR transport in the intergalactic medium (IGM),
i.e., a stochastic, spatially intermittent magnetic field with zero mean
($\langle \bm{B} \rangle = 0$), and under conditions of weak magnetization
($r_g \gg \ell_B$), has not been studied theoretically, numerically, or
experimentally.

\section{Laser-driven experiments} \label{sec:experiments}

\begin{figure}
\begin{center}
\includegraphics[width=\linewidth]{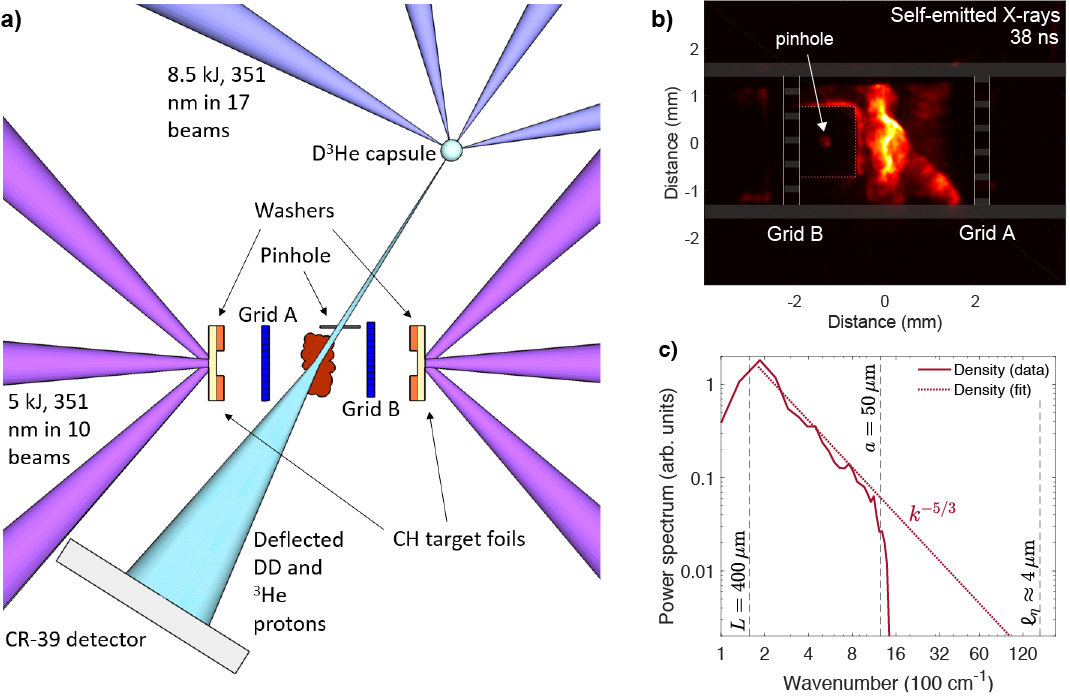}
\caption{\textbf{Experimental setup}.
\textbf{a)} Schematic of the experimental platform, showing the target components
and the configuration of the proton radiography experimental diagnostic. 
\textbf{b)} X-ray self-emission from the interaction region at $t = 38\,\rm{ns}$
after the start of the laser drive. \textbf{c)} Power spectrum of the density
fluctuations  recovered from the fluctuations of X-ray  self-emission. The resolution
of the diagnostic (the size of the pinhole employed on the X-ray framing camera)
is 50 $\mu$m, which is below the driving scale, but above the plasma's dissipative
scales.}
\label{fig:beam_outline}
\end{center}
\end{figure}

Here we report the results of laboratory experiments that focus on this regime.
We carried out these experiments at the Omega Laser Facility at the Laboratory
for Laser Energetics at the University of Rochester \citep{Boehly1997}. A high-velocity,
magnetized, turbulent plasma was generated (Figure~\ref{fig:beam_outline}),
employing the same platform as previously used to demonstrate dynamo amplification of
magnetic fields \citep{Tzeferacos2017pop,Tzeferacos2017}.
Three-dimensional simulations with
the radiation-magnetohydrodynamics code
FLASH \citep{fryxell2000,dubey2009,Tzeferacos2015} guided and informed
the experimental design, including target specifications and the timing of
diagnostics \citep{Tzeferacos2017pop}. 

The platform consists of two 50 $\mu$m-thick polystyrene (CH) foils attached to a
pair of 230 $\mu$m-thick CH washers, with 400 $\mu$m-diameter
machined ``wells'' that act as collimators, placed 8 mm apart. Between the two
targets we position a pair of grids, comprised of periodic
300 $\mu$m holes and 100 $\mu$m wires, placed 4 mm apart. The grid patterns are
shifted to break the mirror symmetry of the system.
Each foil is irradiated with 5 kJ of energy during a 10 ns pulse (10 frequency-tripled
laser beams on each foil staggered in time).
The drive produces two counter-propagating plasma flows, which pass through
the pair of grids, meet,
shear each other, and become turbulent in the central region between the
two grids (the interaction region). For a detailed description of the experimental
platform and the nature of the turbulence it generates, see \cite{Tzeferacos2017}.

\begin{figure}
\centering \includegraphics[width=\linewidth]{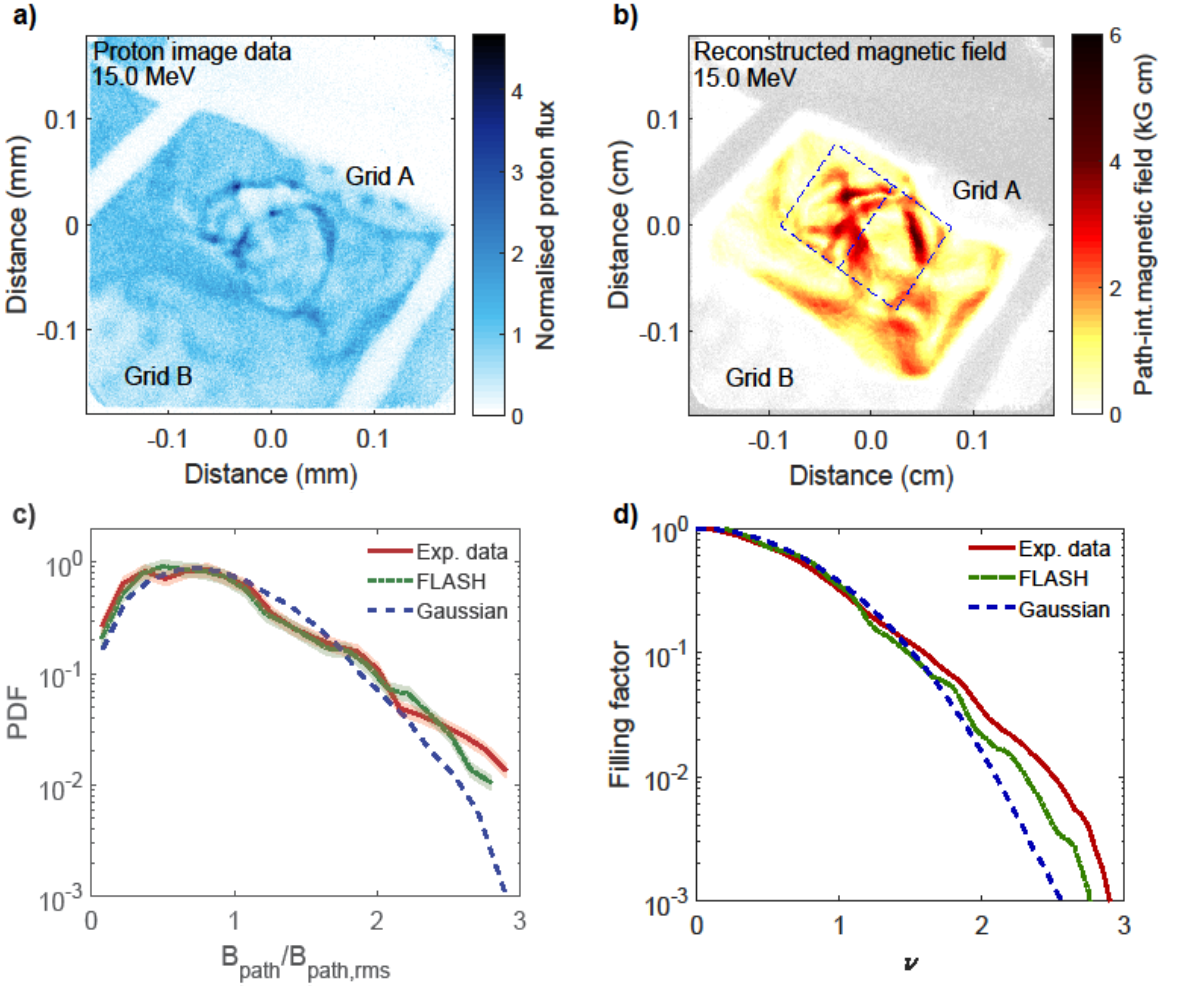}
\caption{\textbf{Magnetic field reconstruction.} \textbf{a)} 15 MeV proton radiography
image of the entire interaction region at 38 ns, without pinhole shield present in the
path. For clarity, the image length scales are shown without the $\times$28
magnification factor; with this factor, the image has dimensions of
10 cm $\times$ 10 cm. \textbf{b)} Magnitude of the two components of path-integrated
magnetic field that are perpendicular to the proton beam path,
reconstructed~\citep{Bott2017} from the 15 MeV proton image in {a)}. \textbf{c)} PDF
of the magnitude of the path-integrated magnetic field $B_{\rm path}$ at 38 ns.
The PDF (red) is calculated using the mean of the PDFs for the two rectangular regions
depicted in {b)}; the uncertainty is derived from the standard error. The PDF of the
path-integrated field arising in the FLASH simulations is also plotted (green),
as is a Gaussian reference (blue) with the same RMS field strength. \textbf{d)}
The fraction of area of the path-integrated magnetic field in which the field's
magnitude $B_{\rm path}$ satisfies $B_{\rm path} \geq \nu B_{\rm path,rms}$,
where $B_{\rm path,rms}$ is the RMS path-integrated field. This quantity is again
calculated from the rectangular regions demarcated in {b)}.}
\label{fig:fullproton}
\end{figure}

The electron density and temperature of the turbulent plasma are measured using
collective Thomson scattering \citep{Katz2012}
and found to be $n_e \simeq 9 \times 10^{19}$ cm$^{-3}$ and $T_e \simeq 400$ eV
immediately after the formation of the turbulent
region ($t \simeq 27$ ns after the start of the drive). The mean velocity
($u_{\rm{flow}}$) and the turbulent velocity ($u_{\rm{turb}}$)
of the flow are also obtained by this diagnostic. Prior to collision, the
counter-propagating flows reach velocities of
$u_{\rm{flow}} \simeq 2 \times 10^7\,\rm{cm}\,\rm{s}^{-1}$, whereas in the
turbulent region at late times we measure
$u_{\rm{flow}} \simeq 5\times10^6 \,\rm{cm}\,\rm{s}^{-1}$ and
$u_{\rm{turb}} \simeq 10^7\,\rm{cm}\,\rm{s}^{-1}$
at the driving scale of the turbulence ($L \simeq 400$ $\mu$m, set by the grid spacing).

The plasma interaction region's evolution is determined using self-emitted
soft X-rays (Figure~\ref{fig:beam_outline}b shows the plasma emission at 38 ns
after the start of the laser drive). As discussed in \citet{Tzeferacos2017},
fluctuations in the emissivity of such a plasma can be related to fluctuations
of density \citep{Churazov2012}; the latter 
exhibit a Kolmogorov power-law spectrum, with driving scale $L$ consistent with
the grid spacing detailed above (Figure~\ref{fig:beam_outline}c). The spatial extent
of the interaction region over time can also be measured using the X-ray diagnostic.
Further details concerning the plasma state are given in Appendix \ref{sec:a.plasma}.

The stochastic magnetic fields amplified in the turbulent plasma
\citep{Tzeferacos2017pop, Tzeferacos2017} are measured using
proton radiography (Figure~\ref{fig:fullproton}). A 420 $\mu$m-diameter SiO$_2$
capsule, with a 2-$\mu$m-thick shell, is filled with 18 atm D$^3$He gas
(6 atm $^2$D and 12 atm $^3$He) and is placed 10 mm away from the interaction region.
The capsule is imploded using 17 beams (frequency-tripled to 351 nm, providing
270 J/beam for a 1 ns pulse) to produce 3.3 and 15 MeV fusion
protons \citep{Li2006, Kugland2012, Manuel2012}. The protons are recorded on the opposite side of the
capsule with a nuclear track detector (CR-39) film pack, 27 cm from the plasma
interaction region, achieving a magnification of $\times$28. Figure~\ref{fig:fullproton}a
shows a proton radiograph of the plasma corresponding to the same time as the X-ray
image in Figure~\ref{fig:beam_outline}b. The presence of strong inhomogeneities in
the proton flux and the stochastic, non-regular morphology of the structures 
is due to protons being deflected by strong, tangled magnetic fields. From the
flux inhomogeneities observed in the proton image, the experimental radiographs can
be inverted \citep{Graziani2017,Bott2017} to recover two components of the
path-integrated magnetic field (Figure~\ref{fig:fullproton}b). Using the measured
spatial extent of the interaction region, it can be shown that the measured
path-integrated magnetic field
corresponds to a root mean square (RMS) value $B_{rms}\simeq 65$-$80$ kG, with a
typical correlation length $\ell_B \approx 90 \, \mu \mathrm{m}$.
These experimental values are consistent with the results of the FLASH simulations
that give $B_{rms}\simeq 80-100$ kG and $\ell_B \simeq 50 \, \mu \mathrm{m}$, when
the effects of diffusion of the imaging beam caused by small-scale magnetic fields
and the underestimation of the magnetic energy by the reconstruction algorithm in
the presence of small-scale caustics are taken into account \citep{Tzeferacos2017}.

The statistics of the path-integrated magnetic field are expected to deviate from
Gaussian as a result of the spatial intermittency. To quantify this, we show the
probability density function (PDF) of the magnitude of the path-integrated field
in Figure~\ref{fig:fullproton}c and the field's filling factor in
Figure~\ref{fig:fullproton}d.  Both exhibit extended tails. Since the
path-integrated magnetic field is spatially intermittent, it follows that the
field itself must also be spatially intermittent. Indeed, the PDF of the
magnetic field strength in the FLASH simulations exhibits an exponential tail.
The simulations also demonstrate that the deviation from Gaussian statistics is
more pronounced in the three-dimensional true fields than the two-dimensional
path-integrated field (see also Appendix \ref{sec:a.plasma}). Such non-Gaussian,
spatially-intermittent magnetic fields are expected to arise when the fluctuation
dynamo is operating~\citep{Schekochihin2004simulations}. 

\section{Transport characterization}\label{sec:char}

\begin{figure}
\centering\includegraphics[width=\linewidth]{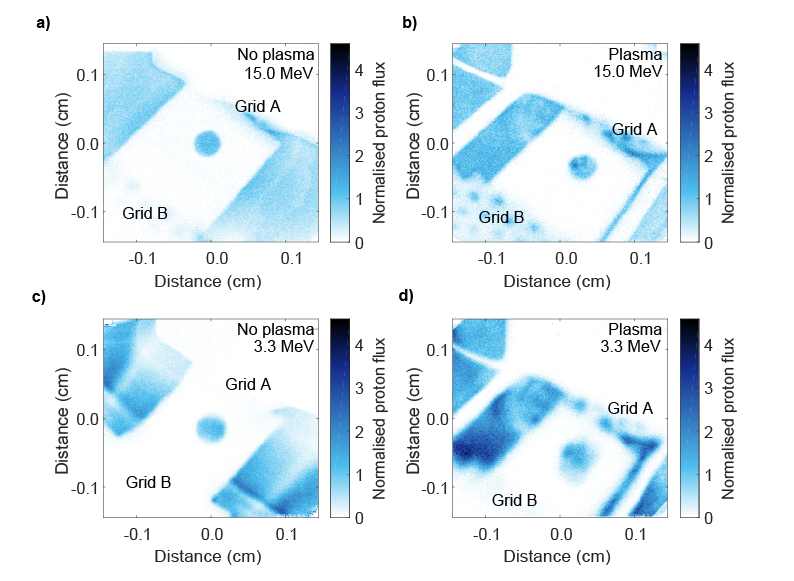}
\caption{\textbf{Proton pinhole images.} Radiographs obtained on the CR-39 film pack
with proton energies of \textbf{a)}
15 MeV, with no plasma in the interaction region, and \textbf{b)} 15 MeV, with a
turbulent plasma in the interaction region.
\textbf{c)} Same as \textbf{a)} but for the 3.3 MeV protons. \textbf{d)} Same as \textbf{b)}
but for the 3.3 MeV protons. 
The pinhole shield is clearly seen to block most of the incoming proton flux from the capsule
and, in the case where no plasma
was present (\textbf{a)} and \textbf{c)}), it produces a fixed 300 $\mu$m diameter beam of 3.3
and 15 MeV protons that passes
through to the detector. For the case when a plasma is present in the interaction region
(\textbf{b)} and \textbf{d)}), the beam
is deformed and broadened before reaching the detector.}
\label{fig:proton_images}
\end{figure}

\begin{figure}
\centering\includegraphics[width=\linewidth]{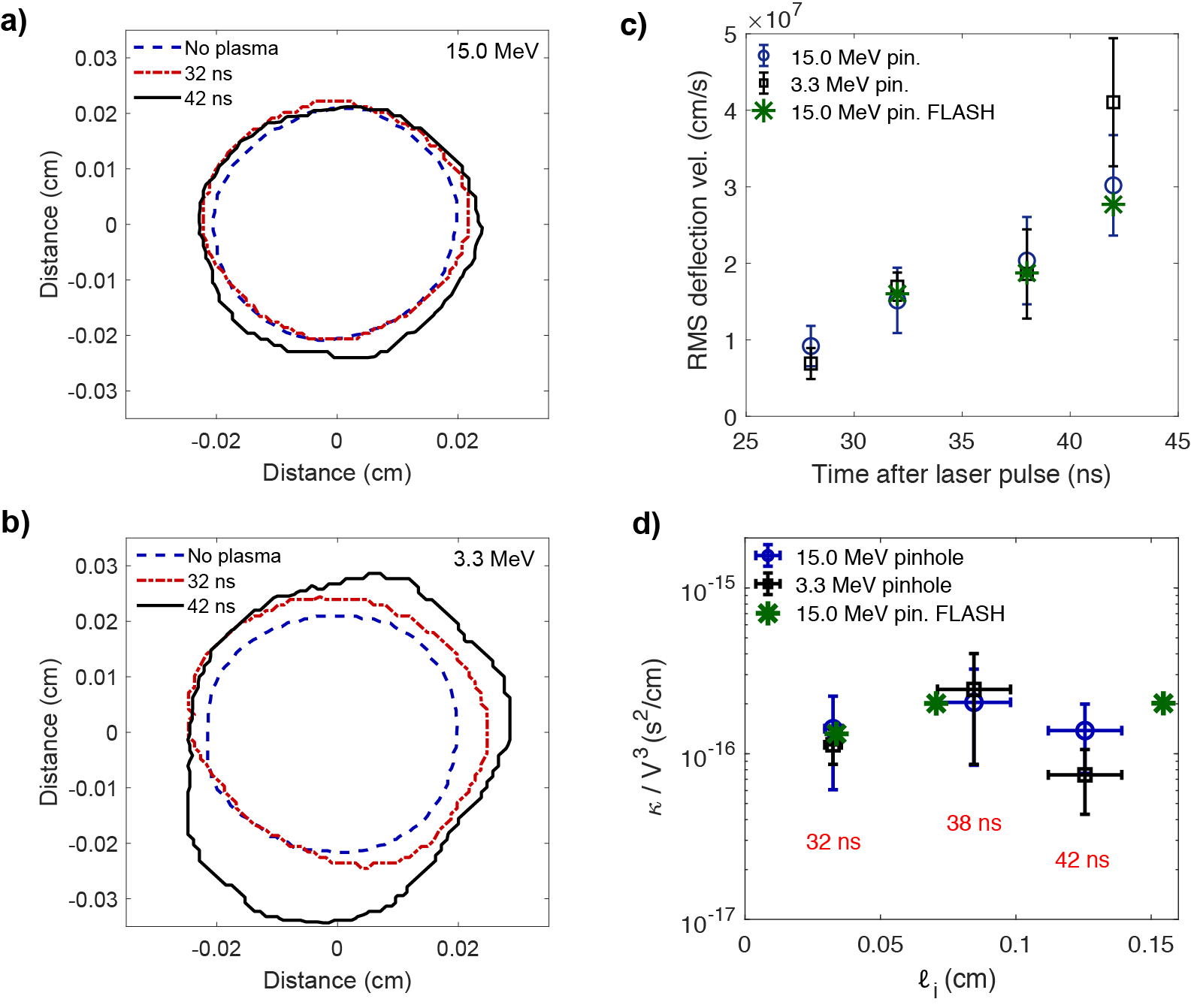}
\caption{\textbf{Diffusive scattering of the proton beam.} Contours of the beam imprint on
the CR-39 plate for \textbf{a)} 15 MeV protons and \textbf{b)} 3.3 MeV protons, taken with
different delay times after the start of the drive beams. \textbf{c)} The RMS transverse
deflection velocity acquired by the proton beam, calculated using the contour analysis of
the pinhole image (for both proton species), and evaluated for the FLASH simulations.
\textbf{d)} Measured spatial diffusion coefficient as a function of the plasma interaction
length, as determined from the X-ray self-emission images and the 15 MeV pinhole synthetic
radiographs from the FLASH simulations.}
\label{fig:lineouts}
\end{figure}

To characterize the transport of particles through the turbulent plasma, we modified
our experimental platform to introduce a collimated
proton beam. The collimation was achieved by placing a 200-$\mu$m-thick aluminum shield
between the D$^3$He capsule and the interaction region,
with a 300-$\mu$m-diameter pinhole (shown in Figure~\ref{fig:beam_outline}a). 
The pinhole imprint is then recorded on the detector plane, as shown in
Figure~\ref{fig:proton_images}.  
The proton-beam imprints appear deformed and broadened due to the interaction
of the protons with the turbulent magnetized plasma.
The proton-beam imprint contours are shown in Figs.~\ref{fig:lineouts}a and \ref{fig:lineouts}b.
The corresponding deflection velocities, $\Delta v_\perp$, as interpreted from the scattering
angle $|\Delta v_\perp|/V$, where $V$ is the proton-beam speed, are shown
in Figure~\ref{fig:lineouts}c.
Using the synthetic proton radiography diagnostic of the FLASH code \citep{Tzeferacos2017pop}
we also post-processed the FLASH simulation results to
recover proton trajectories and the resultant transverse deflections.
These are in good agreement with the experimental measurements
(Figure~\ref{fig:lineouts}c).

The velocity deflection $\Delta v_\perp$ due to magnetic fields scales independently of
velocity, whereas the velocity deflection due to electric fields scales as
$\propto 1/V$  (shown in Appendix \ref{sec:a.deflections}).
The near-equality of the deflection velocities of the two proton species,
evident in Figure~\ref{fig:lineouts}c, suggests that scattering
is predominantly due to magnetic fields.  While there are many other possible processes that
could lead to scattering of a
charged-particle beam passing through a turbulent plasma, for our experiment we argue that
these other processes are negligible, on account of
the low density of the plasma and the large speed of the protons compared to driving-scale
plasma motions. Detailed calculations
and descriptions of possible electric field effects are given in Appendix \ref{a.smearing}.

From our experimental measurement of $\Delta v_\perp$, we can calculate the associated
scattering frequency in velocity-space, 
$\nu \sim(\Delta v_\perp/V)^2 / \tau $, where $\tau = \ell_i/V$ is the transit time of the
particles through the plasma and
$\ell_i$ is the scale of the interaction region as inferred from the X-ray images.
For a plasma with dimensions much bigger than the proton mean free path $\lambda \equiv V/\nu$,
our results imply an isotropic spatial diffusion coefficient $\kappa\sim V^2/\nu =
\ell_i V^3/(\Delta v_\perp)^2$. 
Since $\kappa/V^3$  is constant in our experiment (Figure~\ref{fig:lineouts}d),
this implies $(\Delta v_\perp)^2\propto\ell_i \propto \tau$. This is consistent with a normal
(Markov) spatial diffusion
\citep{Tsytovich1977,ShalchiBook2009,Subedi2017}.

\section{Discussion}\label{sec:discussion}
Since the charged particle transport is consistent with normal spatial diffusion through a
stochastic field,
we can compare the experimental results to theoretical predictions from a random walk process.
Using characteristic values for the plasma properties corresponding to $t = 38$ ns after
the start of the drive, we take the size of the interaction region to be
{\color{black}{$\ell_i \simeq 0.08 \, \mathrm{cm}$}}, the typical magnetic field strength
{\color{black}{$B_{\rm{rms}} \simeq 100 \, \mathrm{kG}$}}, and the correlation length
{\color{black}{$\ell_B \simeq 50 \, \mu \mathrm{m}$}}.
For the case of normal diffusion, a random-walk
argument gives
{$\Delta v_\perp  \approx q_e B_{\rm{rms}} \sqrt{\ell_i \ell_B}/m_p \simeq 1.9 \times 10^7 \, {\rm cm}\,{\rm s}^{-1}$}
(see Appendix \ref{sec:a.deflections}) where $m_p$ is the mass of a proton. This value is consistent
with the measured RMS deflection velocity (Figure~\ref{fig:lineouts}c). Further, since the values
of $V$, $B_{\rm{rms}}$, and the power spectrum of the magnetic energy (and therefore the value
of $\ell_{B}$) do not change in the experiment after the magnetic-field amplification saturates
(see \citet{Tzeferacos2017pop,Tzeferacos2017} and Appendix \ref{sec:a.flash}), the random walk model also predicts
a constant {$\kappa/V^3 \sim m_p^2/(q_e B_{\rm{rms}})^2 \ell_B \simeq 1.9\times10^{-16}\,\rm{s}^{2}\,\rm{cm}^{-1}$},
in quantitative agreement with the experimental results (Figure~\ref{fig:lineouts}d). 

For isotropic statistics and $r_g / \ell_B \gg 1$, the proton mean free path is $\lambda \simeq 10^4\,\rm{cm}$.
In this regime, theory~\citep{Dolginov1967} and simulations~\citep{Subedi2017} predict that 
$\lambda/\ell_B \propto (r_g/\ell_B)^2$. \color{black}{The simulations of \citet{Subedi2017} predict a
scaling coefficient of $1.5$ and extend to $r_g/\ell_B \simeq 40$.  Extrapolating the results of
\citet{Subedi2017} by a factor of $13$ and $28$ to the values $r_g/\ell_B \simeq 520$ for the 3.3 MeV
protons and $\simeq 1,100$ for the 15 MeV protons gives $\lambda/\ell_B\simeq 0.6 \times 10^6$ and $\simeq 2\times 10^6$.
These values agree within a factor of order unity with the experimental value of $\lambda/\ell_B\simeq 2\times10^6$ that
we obtain for the two proton energies. 
More importantly, our results demonstrate that, for the conditions present in the experiment (i.e., a beam with
a diameter $D > \ell_B$ of charged particles in the $r_g \gg \ell_B$ regime that traverses a stochastic and
spatially-intermittent magnetic field with a path length $\ell_i > \ell_B$) the diffusion is not affected
by the spatial intermittency of the stochastic magnetic fields. This is also demonstrated by the FLASH
simulations of the experiment and numerical simulations presented in Appendix \ref{sec:a.intermittency}.

The results of our experiments validate the use of standard diffusion theory in modeling the transport of UHECRs in the IGM, e.g.,
\citet{Kotera2008, Globus2008, Globus2017, Globus2019}, since all three of the above conditions are satisfied. This is useful in
view of the increased interest in such modeling motivated by the recent detection by the Pierre Auger Observatory of a significant
anisotropy in the arrival directions of cosmic rays of energy above 8 EeV \citep{Aab2017a}.

\acknowledgments

The research leading to these results has received funding from the European Research Council under the European Community's Seventh Framework Programme (FP7/2007-2013) / ERC grant agreements no. 256973 and 247039, the U.S. Department of Energy under Contract No. B591485 to Lawrence Livermore National Laboratory, Field Work Proposal No. 57789 to Argonne National Laboratory, and grants no. DE-NA0002724, DE-NA0003605, and DE-SC0016566 to the University of Chicago. We acknowledge support from the National Science Foundation under grants PHY-1619573 and AST-1616037 and from Department of Energy Cooperative Agreement No. DE-NA0001944 to the University of Rochester.   Awards of computer time were provided by the U.S. Department of Energy ASCR Leadership Computing Challenge (ALCC) program. This research used resources of the Argonne Leadership Computing Facility at Argonne National Laboratory, which is supported by the Office of Science of the U.S. Department of Energy under contract DE-AC02-06CH11357. Support from AWE plc., the Engineering and Physical Sciences Research Council (grant numbers EP/M022331/1, EP/N014472/1 and EP/P010059/1) and the Science and Technology Facilities Council of the United Kingdom is also acknowledged, as well as funding from grants 2016R1A5A1013277 and 2017R1A2A1A05071429 of the National Research Foundation of Korea.
J. Matthews, A. Bell and G. Gregori would like to thank Prof. Katherine Blundell (University of Oxford) for stimulating discussions on cosmic-ray measurements.  

\appendix

\section{FLASH simulations}\label{sec:a.flash}

We designed the experimental platform using three dimensional radiation-MHD simulations carried out with the FLASH code. FLASH is a parallel,
multi-physics, adaptive-mesh-refinement (AMR), finite-volume, high performance computing (HPC) Eulerian code \citep{fryxell2000,dubey2009}
that scales well to over a 100,000 processors. This is accomplished by exploiting a variety of in time parallelization techniques, and a combination of
message passing and threading to optimally utilize hardware resources. The code is publicly available (http://flash.uchicago.edu)
and has been successfully applied in a wide range of disciplines including astrophysics, cosmology, combustion, fluid dynamics, turbulence,
and high-energy-density laboratory plasmas (HEDLP).

We performed a series of high-fidelity 3D FLASH radiation-MHD simulations on the Mira supercomputer at the Argonne National
Laboratory. This simulation campaign to design the original experimental platform of  \citet{Tzeferacos2017} is described in detail in \citet{Tzeferacos2017pop};
considerations on the pinhole design are discussed in a companion paper. The simulations take advantage of the entire suite of HEDLP
capabilities of FLASH \citep{Tzeferacos2015}, including its MHD solver \citep{lee2013} extended to three-temperatures \citep{Tzeferacos2015},
non-ideal MHD effects such as magnetic resistivity \citep{Tzeferacos2015} and Biermann battery \citep{fatenejad2013,graziani2015},
heat exchange between ions and electrons, implicit electron thermal conduction and radiation transport in the multi-group diffusion approximation,
multi-temperature tabulated equations of state and material opacities, and laser beams that are modeled using geometric-optics ray-tracing \citep{Kaiser2000}
and deposit energy via inverse Bremsstrahlung.
 
\begin{figure}[htp]
%\begin{center}
\includegraphics[width=\linewidth]{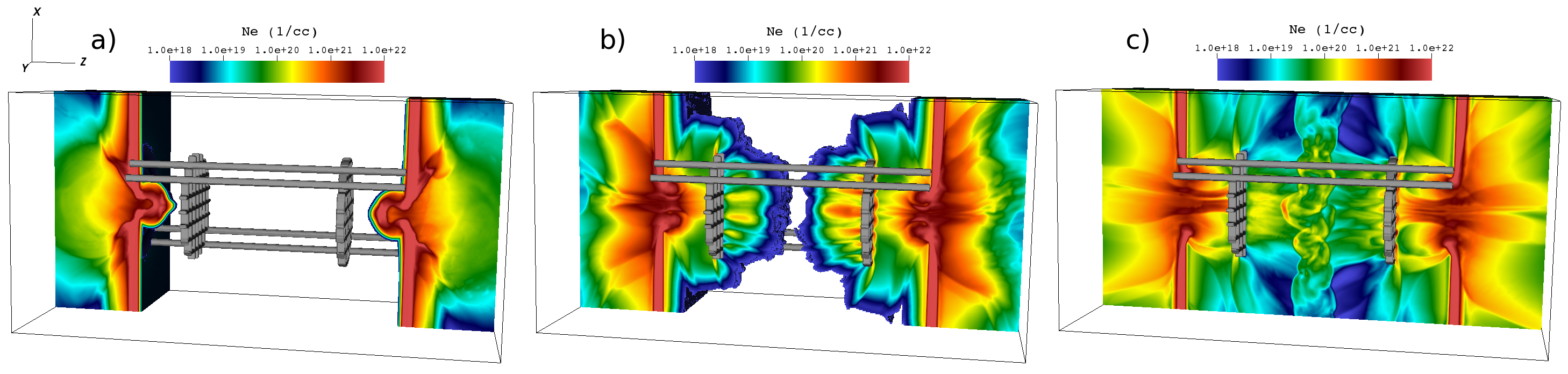}
\caption{\textbf{Simulated temporal evolution}.  We show the logarithm of electron density (half-rendering), with
gray contours denoting the grids and the supporting rods, at \textbf{a)} 11 ns  prior to collision (16 ns in the experiment), \textbf{b)}
1 ns prior to collision (26 ns in the experiment), and \textbf{c)} 16 ns after the collision (42 ns in the experiment) of the counter-propagating plasma flows.}
\label{fig:flash_evolution}
\end{figure}

\begin{figure*}[htp]
\centering
\includegraphics[width=0.5\linewidth]{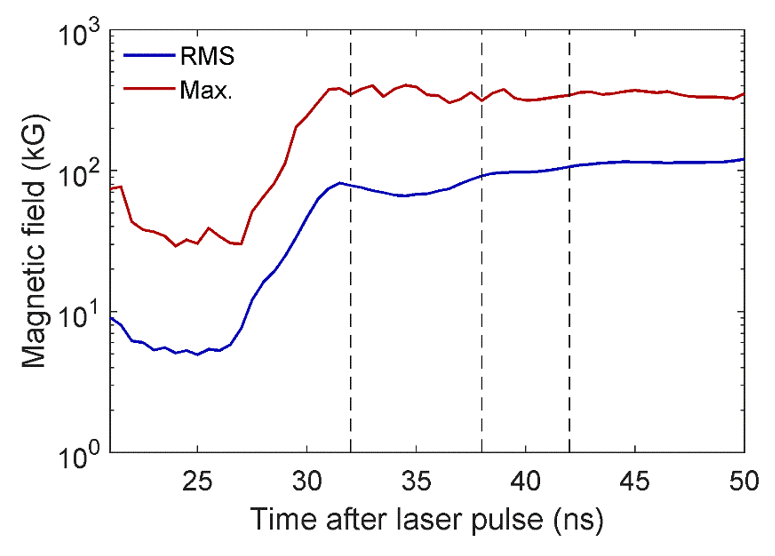}
\caption{\textbf{Temporal evolution of $B_{max}$ and $B_{rms}$}.  Analysis of the FLASH simulations shows that the proton radiography
images were taken (dashed vertical lines) \emph{after} the saturation of the magnetic field amplification.  Thus, the RMS (blue solid line)
and peak (red solid line) values of the magnetic field strength do not vary considerably in time.}
\label{fig:flash_evolutionField}
\end{figure*}

We have used the FLASH code to design and simulate the experiment and, using the code's synthetic diagnostics routines, construct the proton radiographs
shown in Figure \ref{fig:flash_pinhole}.  The latter is used to predict the values of $\Delta v_\perp$ in Figure~\ref{fig:lineouts}c.
The simulation domain spans 0.625 cm in the $X$ and $Y$ directions, and 1.25 cm along $Z$ -- the line of centers between the two targets.
The spatial resolution is $\sim 25$ $\mu$m.  The dynamics of the evolution are shown in Figure \ref{fig:flash_evolution} and largely
mirror the results discussed in \citet{Tzeferacos2017pop,Tzeferacos2017}.  The laser drive ablates the rear surfaces of the  targets to launch two
counter-propagating plasma flows (Figure \ref{fig:flash_evolution}a) that subsequently traverse the grids
(Figure \ref{fig:flash_evolution}b) forming finger-like formations.  The spatial offset of the grids by one grid aperture results
in the two fronts interleaving to create a hot turbulent region in the center of the domain (Figure \ref{fig:flash_evolution}c).
Guided by the simulations, we modified the grids with respect to the design of \citet{Tzeferacos2017} and reduced the thickness of the wires,
from 300 to 100 $\mu$m.  This resulted in an increase in the number of apertures that in turn led to (i) increased throughput of kinetic energy
and to (ii) the formation of a denser interaction region that is considerably thicker and more centered at the target chamber center, when compared
to our previous design in \citet{Tzeferacos2017pop,Tzeferacos2017}.

To quantify the temporal evolution of the magnetic field amplification in the simulations, we utilize a control volume (a cubic box of edge length
500 $\mu$m, \citep{Tzeferacos2017pop}) to initially track in time the propagating plasma front from grid A and, post-collision, the turbulent
interaction region. In the control volume, we compute the peak and RMS magnetic field strengths reported in Figure~\ref{fig:flash_evolutionField}
with red and blue solid lines, respectively. The temporal trend agrees with the results of  \citet{Tzeferacos2017pop,Tzeferacos2017}. The time-series reveals
that, after the amplification saturates, the magnetic field strength does not vary significantly in time and maintains values of $B_{rms}\simeq 80-100\,\rm{kG}$
and $B_{max}\simeq 350-400\,\rm{kG}$, with $\ell_B\simeq50\,\rm{\mu m}$. This corroborates the arguments made Section \ref{sec:char} regarding the scaling of the
spatial diffusion coefficient $\kappa$, since the proton images were taken after saturation (dashed vertical lines in Figure~\ref{fig:flash_evolutionField}).   

\begin{figure}[htp]
  \centering
        \includegraphics[width=0.6\linewidth]{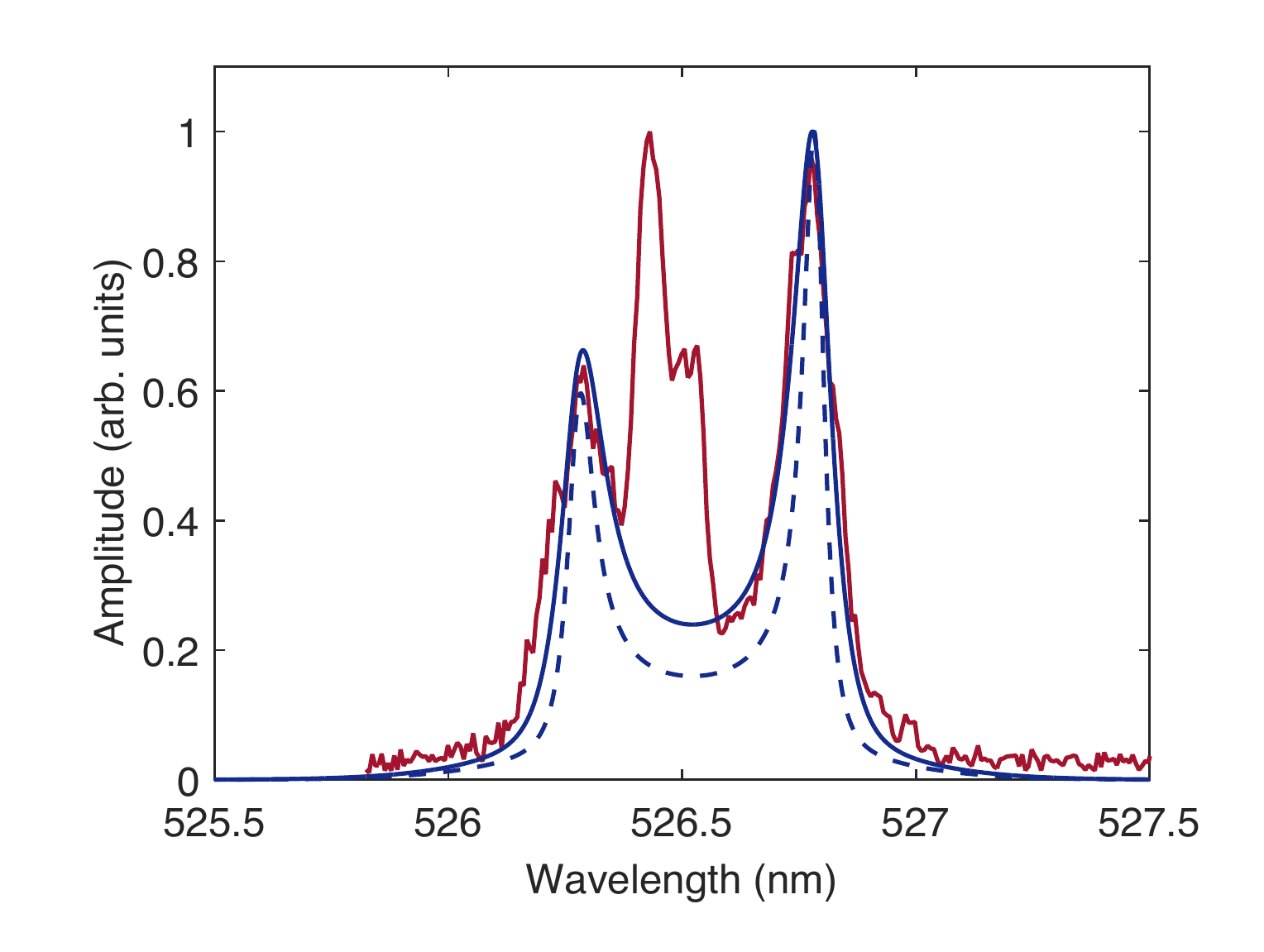}
  \caption{\textbf{Thomson scattering data} The measured ion acoustic wave spectrum at 41 ns after collision is shown in red, and the corresponding
  best fit is shown in solid blue.
Table \protect\ref{tab:TS_data} gives values for the plasma parameters extracted from the fitting procedure.  
The dashed blue line is the best fit determined without inclusion of turbulent broadening.}
\label{fig:TS_example}
\end{figure}

\section{Plasma characterization}\label{sec:a.plasma}
In this Section, we describe the collective Thomson scattering diagnostic which was used to measure plasma properties, and report on measurements taken
just after the collision of the two flows, and much later in time when the turbulent interaction region has developed.  We then provide a comprehensive
list of relevant plasma parameters, both measured and calculated, with particular care in accounting for the multi-ion nature of the plasma. We end this
section by reporting on the evolution of the proton path length and the maximum path-integrated magnetic field strength as a function of time after the
start of the drive laser pulse, which was key in the analysis described in Section \ref{sec:char}.

\subsection{Thomson scattering diagnostic}
A collective Thomson scattering (TS) diagnostic was employed to characterize the plasma properties.  It was operated both as time-resolved or space-resolved
(on different laser shots).  For temporally-resolved TS, the probe beam was a 1 ns pulse, delivering 25 J of 526.5 nm wavelength light to a 50 $\mu \mathrm{m}$ cubed volume.
For spatially-resolved TS, the probe beam was a 0.6 ns pulse, delivering the same energy at the same wavelength to a cylindrical volume with a 50 $\mu \mathrm{m}$ diameter
cross-sectional area, and a 1.5 mm FOV.  For both cases, the scattering angle was set to 63$^\circ$, oriented with the scattering wavenumber parallel to the line of centers
connecting the two targets and therefore the propagation axis of the two plasma flows. Figure~\ref{fig:TS_example} shows an example of the time-resolved measured
data, and the best fit to the spectrum. The central peak at 526.5 ns corresponds to stray light from the probe beam. Fitting the scattering spectra and the total scattering
power with the TS form factor (see~\citet{Tzeferacos2017} for details) allows us to determine the values of the electron density ($n_e$), the electron temperature ($T_e$),
the bulk flow velocity along the propagation axis of the two flows ($u_{flow}$), and the turbulent velocity (i.e., fluctuations which broaden the scattering spectra) at
the scale of the scattering volume ($u_{\ell=50\mu m}$). While the turbulent velocity is measured from TS on scales $\ell = 50\,\mu$m, we can find the turbulent velocity
for the outer scales by assuming Kolmogorov scaling (as shown in Figure \ref{fig:fullproton}d) and see that $u_{turb} \simeq 0.7 - 1.4\times10^7\,\rm{cm}\,\rm{s}^{-1}$. Before
the collision (and until $\sim$ 26 ns) the speed of the 
flow of the bulk plasma is similar to the speed of the single jet, i.e., $u_{jet}  \simeq u_{flow}$, and, naturally, $u_{flow}$ decreases thereafter. These are given in
Table~\ref{tab:TS_data} for 27 ns and 41 ns after the start of the laser pulse.  The 27 ns data was taken using the spatially-resolving configuration,
and the errors in the Table reflect the variation in the plasma parameters across the 1.5 mm extent of the collection cylinder.  The electron density measurement has
an error of about 20-30\%, primarily determined by the uncertainty in the photometric calibration
~\citep{Tzeferacos2017}. It is evident that the average density and temperature in the interaction volume do not vary greatly over the duration of the experiment. 

Plasma conditions derived using collective TS
are complemented by proton radiography measurements, which provide maps of the magnetic field strength \citep{Graziani2017, Bott2017}.  
A comprehensive summary of measured and derived quantities needed to characterize the plasma state fully is given in Table~\ref{tab:PlasmaConds}.
Viscosity and resistivity were calculated following the conventions in \citet{Dorf2014, Helender2002}, respectively. The viscosity depends on the
effective electron-ion collision frequency, which accounts for the presence of multiple ion species present in the plasma. Viscous and resistive scales
were determined according to the conventions given in~\citet{Schekochihin2007a}. The sound speed was calculated for the carbon ion acoustic wave speed,
since this is lower than the hydrogen ion acoustic wave speed.  From this, we see that the turbulence is subsonic.

We see the plasma is collisional and all species are predominantly unmagnetized.
We observe that for conditions occurring at the beginning of the collision between the two flows, the
fluid Reynolds number (Re) and the magnetic Reynolds number (Rm) are of similar magnitude (i.e., the magnetic Prandtl number, Pm, is of order unity).
As the flow evolves, the temperature decreases and Pm gets smaller. Therefore, in general, we expect to be in a low magnetic Prandtl-number regime, ${\rm Pm} \lesssim 1$.

\begin{deluxetable}{cccccc}
\tablenum{1}
\tablecaption{\textbf {Thomson scattering data} Plasma properties computed from two shots at 27 ns and 40 ns after the laser drive.  The spatial/temporal variation respective to each value is reported here for the temperature and velocity measurements, and the electron density measurement has 20-30\% error. \label{tab:TS_data}}
\tablewidth{0pt}
\tablehead{
\colhead{Time after drive} & \colhead{TS mode} & \colhead{$T_{e}$} & \colhead{$n_{e}$} & \colhead{$u_{flow}$}  & \colhead{$u_{\ell=50 \mu m}$}
}
\startdata
27 ns & Spatially-resolved & 400 $\pm$ 80 eV & 9 $\times$ 10$^{19}$ cm$^{-3}$ & $ 1.6\times10^7 \pm 2\times 10^6\,\rm{cm}\,\rm{s}^{-1}$& -- \\
41 ns & Temporally-resolved & 330  $\pm$ 50 eV & 1 $\times$ 10$^{20}$ cm$^{-3}$ & $ 5 \times 10^6 \pm 2 \times 10^6\,\rm{cm}\,\rm{s}^{-1}$ & $3.5-7 \times 10^6\,\rm{cm}\,\rm{s}^{-1}$\\
\enddata
\end{deluxetable}

\begin{deluxetable}{ccc}
\startlongtable
\tablenum{2}
\tablecaption{Summary of measured and calculated plasma parameters.\label{tab:PlasmaConds}}
\tablewidth{0pt}
\tablehead{
\colhead{Plasma parameter} & \colhead{Formula (S.I. units)} & \colhead{Value Time after drive}
}
\startdata
Average atomic weight ($\bar{m}$) & C(50\%), H(50\%)  & 6.5 AMU \\
Temperature ($T$=$T_e$=$T_{i}$) &   & 400 eV \\
Electron density ($n_e$) &   & 9 $\times$ $10^{19}$ cm$^{-3}$ \\
Unit charge ($q$) &   & 1 Electron Charge \\
Carbon mass ($m_C$) &   & 12 AMU \\
Carbon charge ($Z_C$) &   & 6 \\
Hydrogen charge ($Z_H$) &   & 1 \\
Average charge ($\bar{Z}$) &  $\frac{1}{2} \Big( Z_C + Z_H\Big)$ & 3.5 \\
Effective charge ($Z_{eff}$) &  $\frac{1}{n_e} \Big( Z_C^2 n_C + Z_H^2 n_H \Big)$ & 5.3 \\
Ion density ($n_i$) &  $n_e/\bar{Z}$ & 2 $\times$ $10^{19}$ cm$^{-3}$ \\
Respective ion species' densities ($n_H, n_C$)&  $n_H = n_C = \frac{1}{2} n_i$ & 1 $\times$ $10^{19}$ cm$^{-3}$ \\
Electron plasma frequency ($\omega_{pe}$) & $\sqrt{\frac{q^2 n_e}{\epsilon_0 m_e}}$  & $5 \times 10^{14}$ s$^{-1}$ \\
Outer scale ($L$)  &   & $400 \, \mu$m \\
Jet velocity ($u_{jet}$) &  & $\sim 2 \times 10^7\,\rm{cm}\,\rm{s}^{-1}$\\
Turbulent velocity ($u_{turb}$)&   & $\sim 0.7-1.4 \times 10^7\,\rm{cm}\,\rm{s}^{-1}$ \\
Carbon thermal velocity ($v_{th,C}$) & $\sqrt{\frac{2 T}{m_C}}$ &  \\
RMS magnetic field ($B_{rms}$) &   & $\sim 80-100$ kG \\
Debye Length ($\lambda_{D}$)&  $\sqrt{ \frac{T \epsilon_0}{n_{e} q^2} \big(\frac{1}{1 + Z_{eff}}\big)}$ & 5 $\times 10^{-6}$ cm \\
Electron specific heat ratio ($\gamma_e$)&   & 1 \\
Carbon-ion specific heat ratio ($\gamma_C$)&   & 5/3 \\
Carbon-ion sound speed ($c_s$)&  $\sqrt{\frac{\left(Z_C k_B \gamma_e +\gamma_C \right)T}{m_C}}$ & $1.55\times10^7\,\rm{cm}\,\rm{s}^{-1}$ \\
Turbulence Mach number ($\rm Ma$)&  $u_{turb}/c_s$ &  $\sim 0.45-0.95$\\
Plasma $\beta$ & $\frac{\left(n_{e} + n_{i} \right) T}{B_{rms}^2/2 \mu_0}$  & $\sim$ 160 \\
Carbon-carbon impact parameter ($b_{min, C}$) & $ \frac{Z_C^2 q^2}{8 \pi \epsilon_0 T}$ & $5 \times 10^{-9}$ cm \\
Electron-ion impact parameter ($b_{min, e}$) & $\frac{\bar{Z} q^2}{8 \pi \epsilon_0 T}$ & $1 \times 10^{-9}$ cm \\
Coulomb logarithm carbon-carbon ($\Lambda_{CC}$) & log$\Big(\frac{\lambda_{D}}{b_{min, C}}\Big)$   & 6 \\
Coulomb logarithm electron-ion ($\Lambda_{ei}$) & log$\Big(\frac{\lambda_{D}}{b_{min, e}}\Big)$ & 7 \\
Carbon-carbon collision frequency ($\nu_{CC}$) & $4 \pi\frac{Z_C^4 q^4  }{(4 \pi \epsilon_0)^2} \Lambda_{CC} \frac{n_{C}}{m_C^2 v_{th,C}^3} $  & $3 \times 10^{11}$ s$^{-1}$ \\
Electron-electron collision time ($\tau_{ee}$)& $\frac{12 \pi^{3/2}}{2^{1/2}} \frac{n_e q^4 \Lambda_{ei}}{m_e^{1/2} T_e^{3/2} \epsilon_0^2} $   & $4 \times 10^{-12}$ s \\
Effective electron-ion collision frequency ($\nu_{ei, eff}$)& $Z_{eff}/\tau_{ee} $   & $2 \times 10^{12}$ s$^{-1}$ \\
Carbon gyrofrequency ($\Omega_i$)& $\frac{Z_C q B}{m_{C}}$  & $5 \times 10^{8}$ s$^{-1}$ \\
Electron gyrofrequency ($\Omega_e$)&  $ \frac{q B}{m_{e}}$ & $2 \times 10^{12}$ s$^{-1}$ \\
Dynamic viscosity ($\mu$)& $\frac{32 \sqrt{2} }{15 \pi^{3/2}} \frac{\sqrt{m_H}}{Z_C^2 (q/\sqrt{4\pi\epsilon_0})^4} \frac{T^{5/2}}{\Lambda_{ei}}$   & 0.6 $\mathrm{g}\,\mathrm{cm}^{-1}\, \mathrm{s}^{-1}$ \\
Kinematic viscosity ($\nu_c$)& $\mu / \big(n_i \bar{m}\big)$   & 2100 $\mathrm{cm}^{2}\,\mathrm{s}^{-1}$ \\
Reynolds number ($\mathrm{Re}$)& $\frac{u_{turb} L}{\nu} $  & 190 \\
Viscous dissipation scale ($\ell_{\nu_c}$)& $\big(\frac{\nu^3}{\epsilon}\big)^{1/4} \sim \frac{L}{\mathrm{Re}^{3/4}}$  & 8 $\mu$m \\
Spitzer conductivity ($\sigma$)&  $4.1 \frac{n_e q^2}{m_e \nu_{ei,eff}} $ & $87000\,\mathrm{S}\,\mathrm{cm}^{-1}$ \\
Resistivity ($\eta$)&  $\frac{1}{\sigma}$ & $900\,\mathrm{cm}^{2}\,\mathrm{s}^{-1}$ \\
Magnetic Reynolds number ($\rm Rm$)& $\frac{u_{turb} L}{\eta}$  & 440 \\
Magnetic Prandtl number ($\rm Pm$)& $\frac{\rm Rm}{\rm Re}$  & $\sim$ 1 \\
Resistive dissipation scale ($\ell_{\eta}$) &  $\big(\frac{\eta^3}{\epsilon}\big)^{1/4} \sim \frac{\ell_\nu}{\rm Pm^{1/2}}$ & 5 $\mu$m \\
\enddata
\end{deluxetable}

\subsection{Time evolution of plasma}

In Section \ref{sec:char}, we measure the proton spatial diffusion coefficient through a stochastic magnetic field of a particular RMS field strength.
This measurement is taken at different times, corresponding to different extents of plasma expansion and therefore different path lengths. Thus
we require a time-resolved measurement of the path length of the proton beam through the plasma, $\ell_{i}$. This measurement is shown in
Figure~\ref{fig:plasma_charac_time}a.
\begin{figure}[htp]
  \centering
  \includegraphics[width=0.95\linewidth]{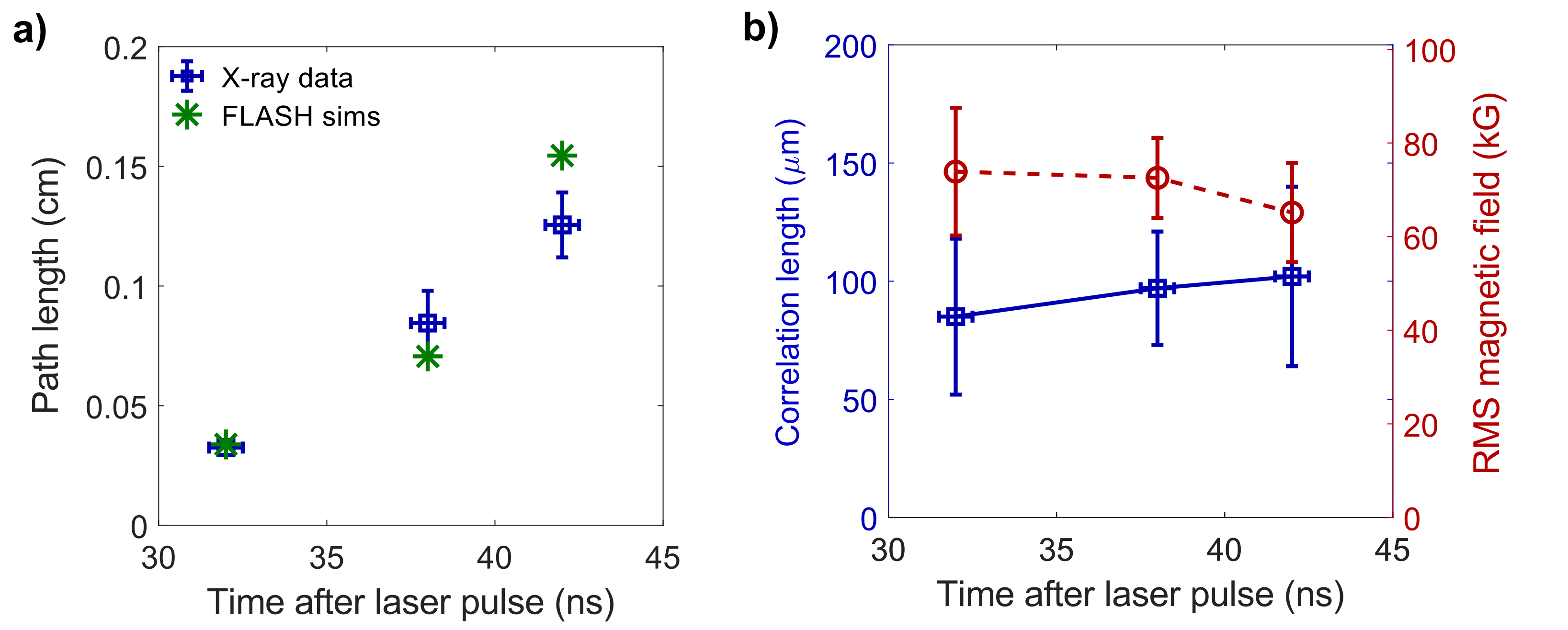}
  \caption{\textbf{Evolution of the interaction region}. \textbf{a)} Plot of the proton-beam path-length over time (shown in blue), as determined
  by measuring the full-half-width-max of the interaction region as seen in X-ray images. For comparison, the proton-beam path-length as directly
  measured in the FLASH simulation is depicted by the green asterisks. \textbf{b)} Plot of the correlation length (blue squares) and RMS magnetic-field
  strength (red circles) of stochastic magnetic fields present in the plasma over time, derived from the reconstructed path-integrated fields. The experimental
  values are consistent with the FLASH simulation values that give $B_{rms}\simeq 80-100$ kG and $\ell_B \simeq 50 \, \mu \mathrm{m}$, when the effects
  of diffusion of the imaging beam caused by small-scale magnetic fields and the underestimation of the magnetic energy by the reconstruction algorithm in
  the presence of small-scale caustics are taken into account \citep{Tzeferacos2017}.}
\label{fig:plasma_charac_time}
\end{figure}

\begin{figure}[htp]
  \centering
  \includegraphics[width=\linewidth]{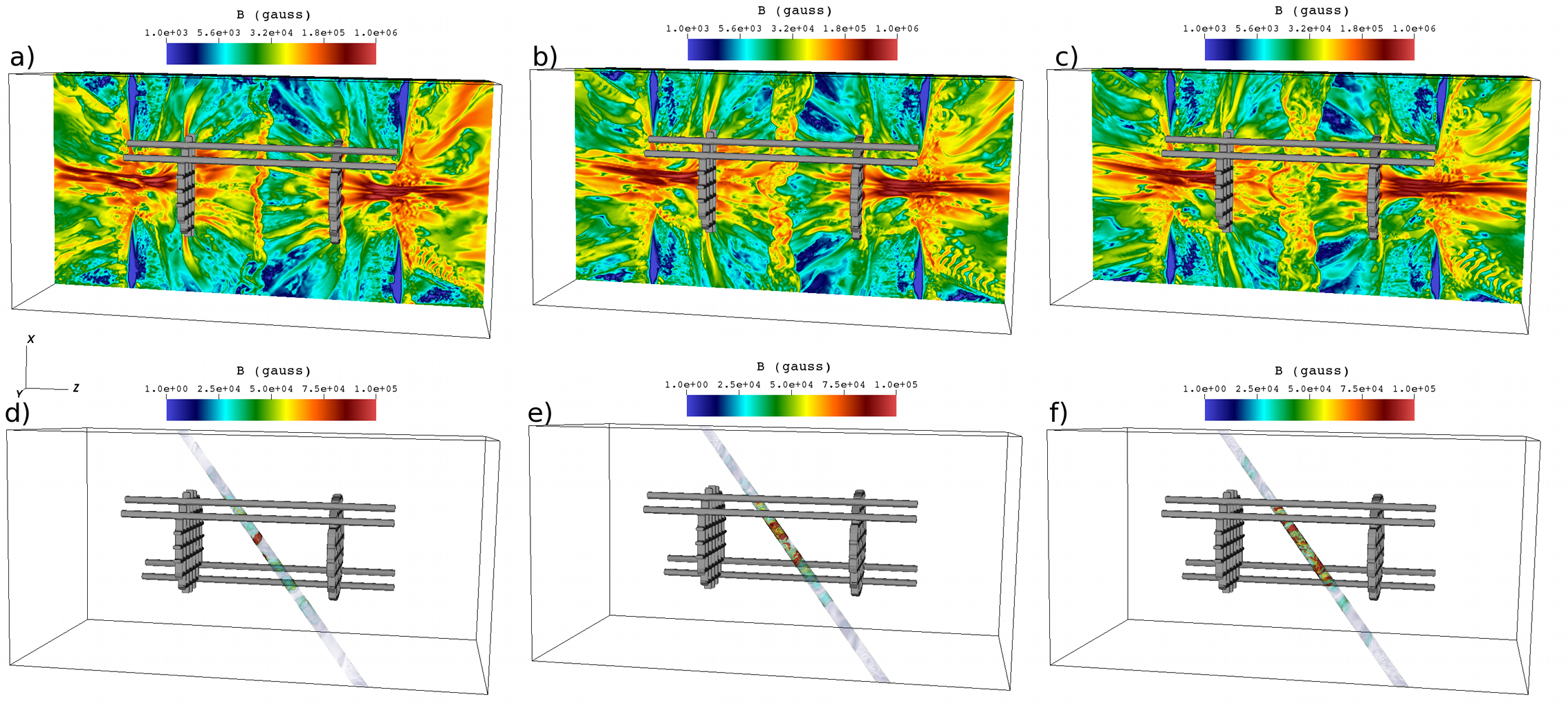}
  \caption{\textbf{FLASH simulations of the magnetic field}. Snapshots of the magnetic-field evolution at times that correspond to \textbf{a)}
  32 ns, \textbf{b)} 38 ns, and \textbf{c)} 42 ns in the experiment.  The intersections of the collimated proton beam with the magnetic field
  of the interaction region is shown for the same times in \textbf{d)-f)}.  We clearly see the expansion of the interaction region with time.}
\label{fig:flash_li}
\end{figure}

We also need to know the RMS field strength and correlation length of the stochastic magnetic field in the plasma; time-resolved measurements of
these quantities are presented in 
Figure~\ref{fig:plasma_charac_time}b.

The measurement of the path length is obtained from the full-width-half-max of the plasma interaction observed in X-ray images. These are obtained
by recording the plasma self-emission onto a framing camera with $\sim$1 ns gate width and filtered with 0.5 $\mu$m $\rm C_2H_4$ and 0.15 $\mu$m Al.
Spatial imaging is achieved using a 50 $\mu$m-diameter pinhole. Thus, the width of the plasma interaction region is defined as being the full-width-half-max
of the X-ray emission profile in the direction parallel to the initial plasma flows. We then calculate the path length by projecting the distance onto
the path of the proton beam through the interaction region (which has a known angle of $55^\circ$ with respect to the direction of the initial plasma flows.
The error in the measurement is determined from the standard deviation of the measured interaction-region width across the image. These measurements are
confirmed by FLASH simulations (see above for a description of the FLASH code and the numerical setup), which show the width of the interaction region
increasing over time: the numerical values obtained directly from the simulation are shown in Figure~\ref{fig:plasma_charac_time}a while
Figure~\ref{fig:flash_li} illustrates the extent of the interaction region at times of 32 ns, 38 ns, and 42 ns.

We derive estimates of the RMS field strength $B_{\rm rms}$ and the correlation length $\ell_B$ of the stochastic magnetic field from the two spatially
resolved components of the path-integrated magnetic field, which are themselves determined from full proton images of the turbulent plasma (see Figure~\ref{fig:fullproton}); the analysis techniques used to perform these calculations are described in~\citet{Bott2017}. To assess the robustness of the result,
we apply the technique to proton images at two additional times ($t = 32, 42$ ns) after the initiation of the laser drive; these images (and the associated
path-integrated fields) are shown in Figure~\ref{fig:prot_images}.
\begin{figure}[htp]
  \centering
  \includegraphics[width=0.8\linewidth]{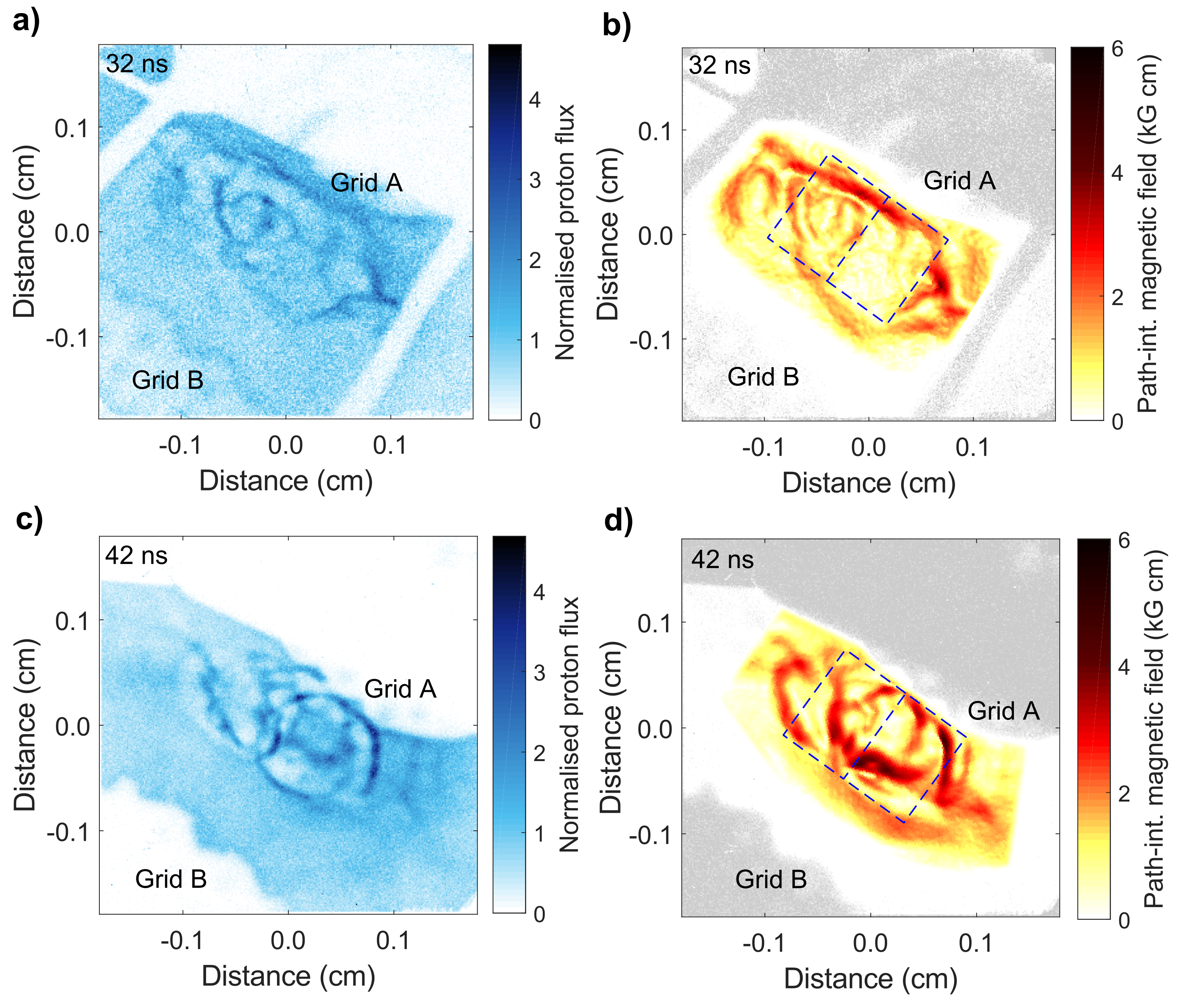}
  \caption{\textbf{Full proton images of the turbulent plasma}. \textbf{a)} Proton image data for $t = 32$ ns after the drive. \textbf{b)}
  Magnitude of perpendicular path-integrated magnetic field at $t = 32$ ns; the regions subsequently analyzed to determine the RMS field strength
  and correlation length are demarcated in a blue dashed line. \textbf{c)} Proton image data for $t = 42$ ns after the drive. \textbf{d)} Magnitude
  of perpendicular path-integrated magnetic field at $t = 42$ ns.}
\label{fig:prot_images}
\end{figure}
To derive an estimate of the uncertainty on the measurements, we perform these calculations for two rectangular regions, and use the means and errors
to determine the results plotted in Figure~\ref{fig:plasma_charac_time}b. We find that within the experimental uncertainty of the measurement,
the RMS field strength and correlation length does not change in time. Further, the experimental values are consistent with the FLASH simulation values
that give $B_{rms}\simeq 80-100$ kG and $\ell_B \simeq 50 \, \mu \mathrm{m}$, when the effects of diffusion of the imaging beam caused by small-scale
magnetic fields and the underestimation of the magnetic energy by the reconstruction algorithm in the presence of small-scale caustics are taken
into account \citep{Tzeferacos2017}. 

\subsection{Magnetic-field spatial intermittency}

In Section \ref{sec:experiments} , we state that the observed spatial intermittency of the measured components of path-integrated magnetic field provided evidence that
the stochastic fields themselves were spatially-intermittent; here, we demonstrate this claim explicitly for the FLASH simulations.
Figure~\ref{fig:flash_intermittency}a and Figure~\ref{fig:flash_intermittency}b show the PDFs and filling factors respectively
of the true three-dimensional field, along with the equivalent quantities for a Gaussian stochastic field. 
\begin{figure}[htp]
  \centering
  \includegraphics[width=0.9\linewidth]{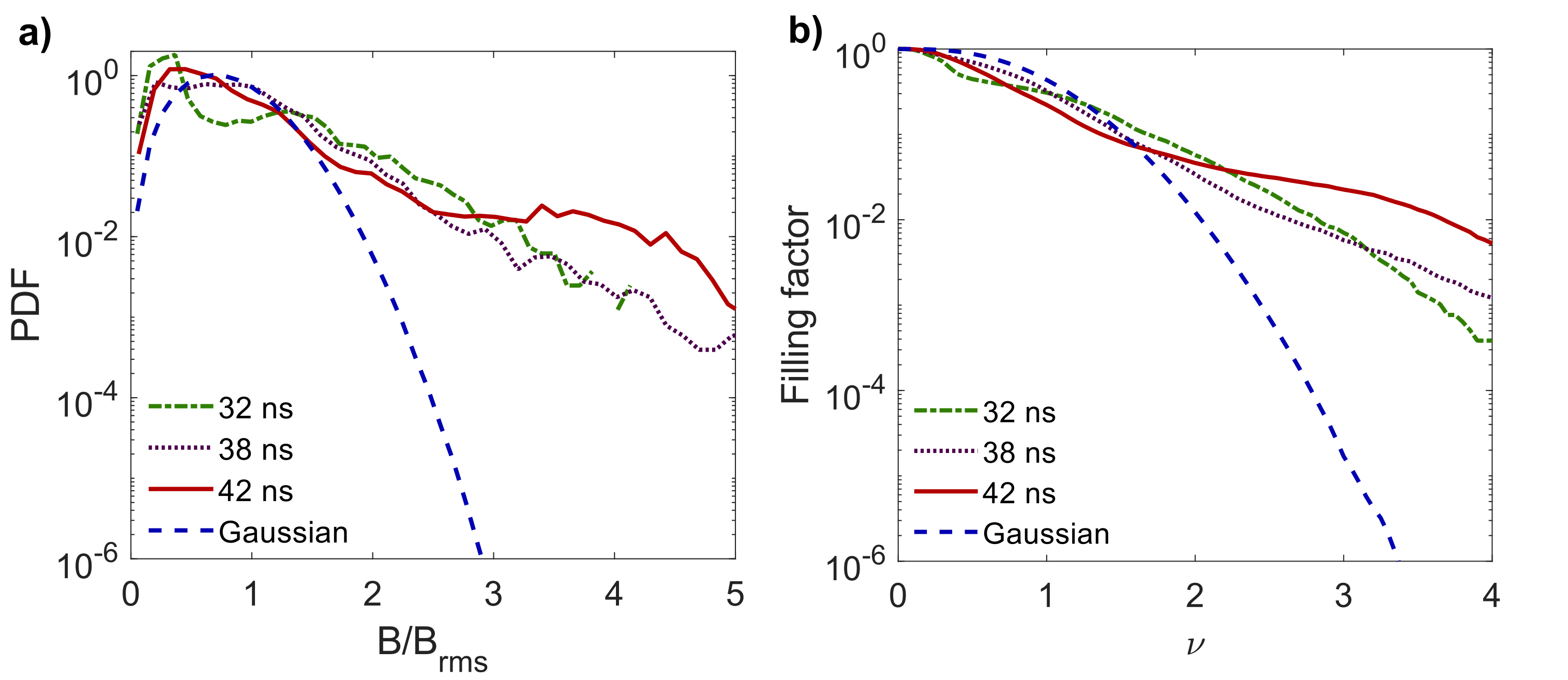}
  \caption{\textbf{Spatial Intermittency of FLASH-simulated magnetic fields}.\textbf{a)} Probability-density function (PDF) of the magnetic fields
  contained inside the sections of the cylindrical volumes depicted in Figures~\ref{fig:flash_li}d, \ref{fig:flash_li}e,
  and \ref{fig:flash_li}f which are also contained within the full-half-width-maximum of the field with respect to the proton path length.
  A Gaussian with unit RMS is plotted for reference. \textbf{b)} The filling factor (see caption of Figure \ref{fig:fullproton}) for the same
  regions of the stochastic magnetic field.}
\label{fig:flash_intermittency}
\end{figure}

\begin{figure}[htp]
  \centering
  \includegraphics[width=0.9\linewidth]{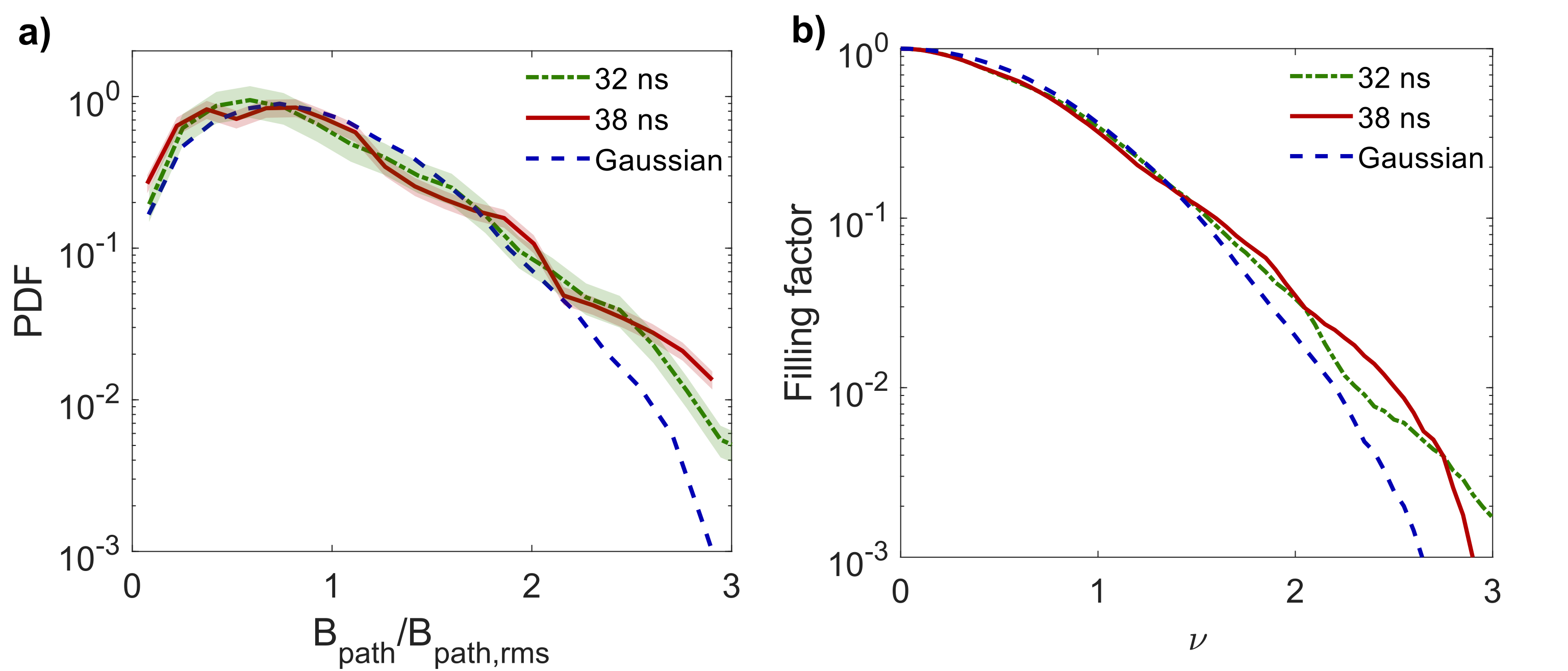}
  \caption{\textbf{Time evolution of path-integrated field spatial intermittency}. \textbf{a)} The PDF of the magnitude of the (perpendicular)
  path-integrated magnetic field measured using the full proton-imaging diagnostic at 32 ns and 38 ns. These quantities are calculated for the
  rectangular regions denoted in Figure~\ref{fig:prot_images}b for $t = 32$ns, and Figure~\ref{fig:fullproton} for $t = 38$ns. As
  in Section \ref{sec:experiments}, a reference PDF for a Gaussian path-integrated field is plotted. \textbf{b)} The filling factor for the magnitude of the
  path-integrated field at 32 ns and 38 ns, along with the filling factor for a Gaussian random field.}
\label{fig:data_intermittency}
\end{figure}

We observe that at all times in the simulation the departure from Gaussian behavior is much more significant than the equivalent departure
observed for the path-integrated field. In the simulations, for $B/B_{\rm rms} > 1$, an exponential fit is obtained for the PDF; this matches
the results of previous simulations of the turbulent dynamo~\citep{Schekochihin2004simulations}.

To assess the robustness of our results, we also consider the PDF and filling factor of the experimentally recovered components of the
path-integrated at 32 ns, and compare them to the equivalent quantities (at 38 ns) depicted in Section \ref{sec:experiments}; the results are shown in
Figure~\ref{fig:data_intermittency}. 

We find in both cases that there is a departure from Gaussian behaviour at sufficiently large values of the path-integrated field magnitude
$B_{\rm path}$ as compared to the RMS value $B_{\rm path,rms}$, and also for sufficiently large values of the filling factor $\nu$. However,
we also observe some fluctuation in the precise shape of both the PDF and the filling factor, which in turn suggests that such fluctuations
should not be regarded as being physically significant. 
 
\section{Detailed calculations of proton deflections}\label{sec:a.deflections}
In this Section, we discuss how the deflection velocity due to scattering angle, $\Delta v_\perp$, of the proton beam is related to the
stochastic magnetic and electric fields present in the plasma. We also demonstrate through numerical tests that the excess blurring in
the experimental proton image compared to the synthetic images from the reconstruction algorithm 
is due to diffusive scattering by small-scale magnetic field structures (see Figure \ref{fig:fullproton}). We then describe the procedures
used to provide quantitative estimates of $\Delta v_\perp$ shown in Figure \ref{fig:lineouts}c from both the magnetic-
field reconstruction technique paired with the smearing analysis of the driving-scale structures (illustrated in Figure \ref{fig:fullproton}),
and from the pinhole-edge contour analysis from the pinhole proton radiography (Figures \ref{fig:lineouts}a and \ref{fig:lineouts}b). 
  
\subsection{Calculation of diffusive deflection velocities for stochastic electromagnetic fields}

The change in transverse velocity $\Delta \bm{v}_B$ due to angular scattering of a proton of initial speed $V$ due to the presence of a
magnetic field can be approximated in the small-deflections limit, $|\Delta \bm{v}_B| \ll V$, by the following procedure: We begin by noting
that the length of time the particle experiences the magnetic-field is given by $\tau \approx z/V$, where $z$ is the path length. Making this
substitution and integrating the proton's equation of motion (with $m_p$ the proton mass and $q$ its charge), we find 
\begin{equation}
\Delta \bm{v}_B \approx \frac{q}{m_p}\bm{\hat{z}} \times \int_{0}^{\ell_i} \bm{B_{\perp}} \, \mathrm{d}z \, ,
\label{eqn:Bintegral}
\end{equation}
where $\hat{z}$ is the unit vector of $z$ and $B_{\perp}$ is the perpendicular component of the magnetic field relative to $z$.  We
see $\Delta v_B$ has no dependence on the initial velocity. The deflection angle is
\begin{equation}
\delta \theta_B = \frac{|\Delta \bm{v}_B|}{V} \, ,
\end{equation}
and the deflection length, $\delta_B$, which is the displacement of the proton from its original position due to deflection by the magnetic field, is
\begin{equation}
\delta_B = r_{det}\delta \theta_B \propto \frac{1}{V} \, ,
\end{equation}
where $r_{det}$ is the distance from where this deflection occured to the detector.  Similarly, for electric fields, we have
\begin{equation}
\Delta \bm{v}_E = \frac{q}{m_p V}
\int_{0}^{\ell_i} \bm{E}_{\perp} \, \mathrm{d}z \propto \frac{1}{V} \, , \label{eqn:Eintegral}
\end{equation}
and the deflection angle is 
\begin{equation}
\delta \theta_E = \frac{|\Delta \bm{v}_E|}{V^2} \, .
\end{equation}
The deflection length due to electric fields is, therefore,
\begin{equation}
\delta_E = r_{det}\delta \theta_E \propto \frac{1}{V^2}.
\end{equation}

Having derived $\Delta v_\perp$ due to arbitrary magnetic and/or electric fields, we specialize to stochastic fields. An expression for
$\Delta v_\perp$ as a function of the RMS magnetic field strength $B_{rms}$ and the field's correlation scale $\ell_B$ -- \color{black}{which,
on account of being the characteristic scale over which the magnetic field falls to zero from its peak value, is approximately half the size
of the typical magnetic structure}\color{black} -- can be derived by the following heuristic argument. From Equation \ref{eqn:Bintegral}, the
change in velocity acquired by a proton traversing a single magnetic structure of size $2 \ell_B$ and strength $B_{rms}$ is given by
\begin{equation}
\delta v_{\ell_B} \sim \frac{2 q B_{rms} \ell_B}{m_p}.
\end{equation}
Assuming the velocity deflections add as a random walk, the overall velocity deflection can be approximated by
$\Delta v_B \simeq \delta v_{\ell_B} \sqrt{N_0}$, where $N_0 \gg 1$ is the number of structures encountered by the proton.
Since $N_0 \approx \ell_i/2 \ell_B$, where $\ell_i$ is the path length of the proton traversing the interaction region, it
follows that $\Delta v_B \sim \delta v_{\ell_B} \sqrt{\ell_i/2 \ell_B}$, leading to
\begin{equation}
\Delta v_B \approx \frac{q B_{rms} \sqrt{\ell_i \ell_B}}{m_p} \, , \label{eq:deltavB}
\end{equation}
where we have adjusted the numerical pre-factor to agree with a rigorous correlation analysis~\citep{Bott2017}
We note that $\Delta v_B$ is independent of the initial proton velocity $V$, indicating that both proton species will
experience the \textit{same} diffusive deflection velocity due to scattering by the magnetic field.  The deflection angle due to the magnetic field is then
\begin{equation}
\delta \theta_B \approx \frac{\Delta v_B}{V} \approx \frac{q B_{rms}\sqrt{\ell_i \ell_B}}{m_p V} \, ,
\end{equation}
and the deflection length is
\begin{equation}
\delta_B \approx r_{det} \frac{q B_{rms} \sqrt{\ell_i \ell_B}}{m_p V} \, .
\label{eq:deltaB}
\end{equation}
For a stochastic electric field with RMS field strength of $E_{rms}$ and scale of $\ell_E$, an expression similar to Equation \ref{eq:deltavB} can be found:
\begin{equation}
\Delta v_E \approx \frac{q E_{rms} \sqrt{\ell_i \ell_E}}{m_p V} \, . \label{eq:deltavE}
\end{equation}
We note that since $\Delta v_E \propto 1/V$, faster protons will experience a smaller diffusive deflection velocity due to electric field scattering.

The total deflection velocity due to scattering angle will have contributions from both magnetic and electric fields,
\begin{equation}
\Delta v_\perp = \Delta v_B + \Delta v_E \, .
\end{equation}
For a single proton, it is impossible to distinguish between deflections due to magnetic and electric fields. However, in our experiment we
generate protons of distinct energies (15.0 MeV protons from $\rm D^3He$ nuclear reactions, and 3.3 MeV protons from DD reactions). These
protons experience the same fields - we now show that the relevant timescales allow this. The eddy-turnover time at the driving scale
is $L/u_{turb}\simeq$ 4 ns, while the shortest turnover time of plasma motions occurs at the dissipation scale $\ell_{\eta}$.  Following
the Kolmogorov scaling, $u_{turb,\ell_{\eta}} \sim u_{turb} (\ell_{\eta}/L)^{1/3}$, which yields $\ell_{\eta}/u_{turb,\ell_{\eta}}\simeq$ 0.2 ns,
which is larger than the transit time of the proton beams across the plasma ($\tau$), the pulse length of the beams ($\tau_p$), and the time
difference between the transit of the proton beams ($\tau_d$). Specifically, for the 3.3 MeV DD protons,
$\tau \approx \ell_i/V_{\mathrm{DD}} \approx 40 \, \mathrm{ps}$, $\tau_p \approx  150 \, \mathrm{ps}$, $\tau_d \approx r_i/V_{\mathrm{DD}} -r_i/V_{\mathrm{D^3He}} \approx 100 \, \mathrm{ps}$
($r_i$ is the distance from the proton source to the plasma). Thus, both proton beams sample the same fields, and so we can use the different velocity scalings
illustrated by Equations \ref{eqn:Bintegral} and \ref{eqn:Eintegral} to distinguish between $\Delta v_B$ and $\Delta v_E$.

\subsection{Experimental measurement of diffusive deflection velocity from proton pinhole-edge contours}

For the calculation of the pinhole-edge contours for both 15.0 MeV and 3.3 MeV proton shown in Figures \ref{fig:lineouts}a and \ref{fig:lineouts}b, we
adopt a three-step procedure. First, the center of proton flux of the pinhole is calculated, using the flux distribution as a weight function.
In principle, the presence of driving-scale inhomogeneities could mean that this is different to the projected center of the pinhole beam, given
the position of the pinhole; however, for our data the difference between these points is negligibly small compared to the radius of the pinhole
$R = 150 \, \mu \mathrm{m}$. We then calculate the mean proton flux inside the specified pinhole radius, $\Psi_0$ (taking account of the magnification).
Finally, we define the contour of interest as the contour corresponding to 20\% relative to the mean proton flux. 

The motivation of this definition for pinhole-edge contour arises from two considerations: one theoretical, and one practical. Theoretically, deflections
of the proton beam can be due to both driving-scale and small-scale structures.
The former result in coherent distortions of the pinhole-edge contour from its undeflected position; the latter leads to a general expansion
of the pinhole contour, due to a shallowing of the gradient of the pinhole flux distribution at its edge. The 20\% contour is sensitive to
both effects (unlike the 50\% contour, which is not sensitive to the small-scale structures). Practically, due to its large extent, the 20\% contour
is large enough to provide a reasonable average over the driving-scale structures. We do not choose an even smaller percentage value for the contour
because of distortions to such contours arising from noise in the CR-39 plate in regions where the pinhole shield is present.  

To extract quantitative estimates of $\Delta v_{\bot}$, the typical deflection velocity for a given proton beam energy from the pinhole-edge contours,
we compare the average deviation of the pinhole-edge contours obtained in the presence of the turbulent plasma with the calibration pinhole-edge contour
for the same energy (i.e., with no plasma present). The comparison is performed numerically by measuring the squared displacement
$\Delta r^2(\theta) \equiv [r_{plas}(\theta) - r_{cal}(\theta)]^2$ between the plasma pinhole-edge contour radius $r_{plas}(\theta)$ and the calibration
plasma pinhole-edge contour radius $r_{cal}(\theta)$ along a ray with polar angle $\theta$ originating from the centre of proton flux. The averaged
displacement is then
\begin{equation}
\Delta r_{rms} \equiv \left[\frac{1}{2 \pi} \int_0^{2 \pi} \mathrm{d} \theta \, \Delta r^2(\theta)\right]^{1/2} \, . 
\end{equation}
The error in the displacement is in turn calculated from the variance of this measure. We find that small changes in the percentage value of the
pinhole-edge contour do not significantly change our result. 

We move from $\Delta r_{rms}$ to a measurement of the typical deflection velocity using the fact that, for any small-angle  scattering process
with typical deflection magnitude $\Delta v_\perp$, and associated deflection length $\delta \equiv r_{det} \Delta v_\perp/V$, the effect of the
diffusive scattering on the proton flux distribution can by modeled quantitatively by~\citet{Bott2017}
\begin{equation}
\Psi(\bm{x}_{\bot})= \int \mathrm{d}^2 \bm{\widetilde{x}}_{\bot} \,  \widetilde{\Psi}(\bm{\widetilde{x}}_{\bot})\frac{1}{\pi \delta^2} \mathrm{exp} \Big[-\Big(\frac{\bm{x}_{\bot}-\bm{\widetilde{x}}_{\bot}}{\delta} \Big)^2 \Big]
\label{eqn:fluxdistribution} \, ,
\end{equation}
where $\bm{x}_{\bot}$ denotes perpendicular position on the detector, $\bm{\widetilde{x}}_{\bot}$ is a (two-dimensional) integration variable
representing positions near $\bm{x}_{\bot}$, $\Psi(\bm{x}_{\bot})$ is the proton flux distribution at $\bm{x}_{\bot}$, and $\widetilde{\Psi}(\bm{x}_{\bot})$
is the  proton flux distribution in the absence of the diffusive scattering process (evaluated at integration variable $\bm{\widetilde{x}}_{\bot}$). 
The model can be simplified under the assumption that the initial pinhole flux distribution is a uniform distribution, with with radius $R$ and
mean flux $\Psi_0$. By symmetry, the smeared pinhole proton flux distribution $\Psi(\bm{x}_{\bot})$ is only a function of radial distance $r$ from
pinhole center-of-mass under this assumption; that is, $\Psi(\bm{x}_{\bot})= \Psi(r)$. Introducing polar coordinates $(\tilde{r},\tilde{\theta})$
for integration variable $\bm{\widetilde{x}}_{\bot}$, where $\tilde{\theta}$ denotes the angle between vectors $\bm{x}_{\bot}$ and $\bm{\widetilde{x}}_{\bot}$,
Equation \ref{eqn:fluxdistribution} becomes
\begin{equation}
\Psi(r)= \frac{\Psi_0}{\pi \delta^2} \mathrm{exp} \Big[-\Big(\frac{r}{\delta} \Big)^2 \Big] \int_0^{R} \mathrm{d} \tilde{r} \, \tilde{r} \mathrm{exp} \Big[-\Big(\frac{\tilde{r}}{\delta} \Big)^2 \Big]
\int_{-\pi}^{\pi}  \mathrm{d} \tilde{\theta} \, \mathrm{exp} \Big[\frac{2 r \tilde{r} \cos\theta}{\delta^2}\Big] \, .  \label{eqn:fluxdistribution_polar}
\end{equation}
Under the further assumption that the radius of the pinhole $R$ is much greater than the smearing parameter $\delta/\mathcal{M}$ (normalized by the magnification factor),
we have that for values of the radial distance $r \sim R$, the dominant contribution to the integral in Equation \ref{eqn:fluxdistribution_polar} arises
for $\tilde{r} \sim r$, and $\theta \ll 1$. These constraints give
\begin{equation}
\int_{-\pi}^{\pi}  \mathrm{d} \tilde{\theta} \, \mathrm{exp} \Big[\frac{2 r \tilde{r} \cos\theta}{\delta^2}\Big] \approx \frac{\sqrt{\pi} \delta}{\sqrt{r \tilde{r}}} \, ,
\end{equation}
leaving
\begin{equation}
\Psi(r) \approx \frac{\Psi_0}{\sqrt{\pi} \delta} \int_0^{R} \mathrm{d} \tilde{r} \,  \sqrt{\frac{\tilde{r}}{r}} \mathrm{exp} \Big[-\Big(\frac{r-\tilde{r}}{\delta} \Big)^2 \Big]
\label{eqn:fluxdistribution_1D_A} \, .
\end{equation}
Since the dominant contribution to the integral with respect to $\tilde{r}$ comes from $\tilde{r}$ satisfying $\tilde{r} -r \sim \delta$, we
can approximate $\tilde{r} \approx r$; Equation \ref{eqn:fluxdistribution_1D_A} then simplifies to
\begin{eqnarray}
\Psi(r) & \approx & \frac{\Psi_0}{\sqrt{\pi} \delta} \int_{0}^{R} \mathrm{d} \tilde{r} \, \mathrm{exp} \Big[-\Big(\frac{r-\tilde{r}}{\delta} \Big)^2 \Big] \nonumber \\
& \approx & \frac{\Psi_0}{2} \mathrm{ erfc} \Big( \frac{r-R}{\delta} \Big)
\label{eqn:fluxdistribution_1D_B} \, ,
\end{eqnarray}
for $\mathrm{ erfc}(x)$ the complementary error function. We conclude that under this model, the 20\% contour on average will satisfy 
\begin{equation}
\frac{\Psi_0}{5} \approx \frac{\Psi_0}{2} \mathrm{ erfc} \Big( \frac{\Delta r_{rms}}{\delta_{rms}} \Big) \, ,
\end{equation}
where $\delta_{rms}$ is the RMS deflection length. It then follows that
\begin{equation}
\delta_{rms} \approx \Delta r_{rms}/\mathrm{ erfc}^{-1}\left(\!\frac{2}{5}\right) \, \approx 1.65 \Delta r_{rms}.
\end{equation}

\begin{figure}[htp]
  \centering
  \includegraphics[width=0.87\linewidth]{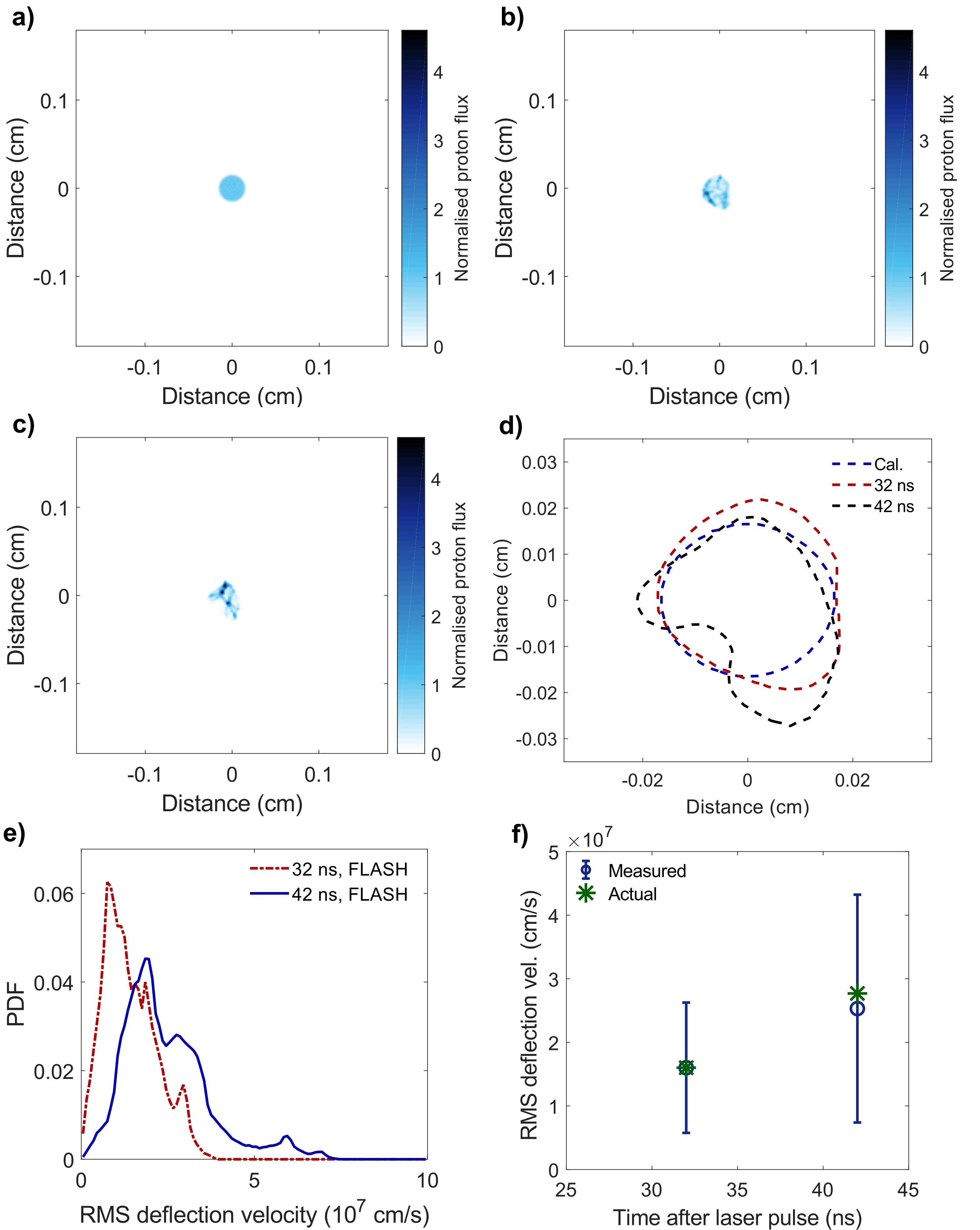}
  \caption{\textbf{Pinhole projections generated by FLASH simulations.} Synthetic images of a collimated proton beam transiting through a
  turbulent plasma produced using FLASH.  These correspond to \textbf{a)} a calibration image with no plasma present and delay times of \textbf{b)}
  32 ns and \textbf{c)} 42 ns.  \textbf{d)} The contours of these pinhole shapes were taken, analogous to the contours in Figure \ref{fig:lineouts}a and b. \textbf{e)} Probability density function (PDF) of simulated proton beam passing through the FLASH simulation at 32 ns
  and 42 ns. \textbf{f)} Estimate of RMS deflection velocity as calculated from the FLASH pinhole-proton edge contours for 32 ns and 42 ns, as
  compared to the true values.} \label{fig:flash_pinhole}
\end{figure}

\subsection{FLASH simulations of proton diffusion}
FLASH simulations are used to create synthetic images of the proton flux, through a 300 $\mu$m diameter pinhole, recorded on the CR-39 plate. This is shown in
Figure \ref{fig:flash_pinhole}a, b, and c for different probing times. The collimated beam consists of 500,000 simulated protons.
The resolution of the FLASH synthetic radiographs, determined by the imposed capsule smearing and the proton binning, is 50 $\mu$m to match the
experimental resolution.
Next, we evaluate the 20\% pinhole-edge contour in the same manner we have applied to the experimental data, and show the results in Supplementary
Figure \ref{fig:flash_pinhole}d. Notably, the profile at 42 ns after the laser drive exhibits significant anisotropy.  This is naturally expected --
in probing a relatively small area in a stochastic environment, it is reasonable that we would at some point probe an anisotropic portion in the magnetic
field distribution.  In fact, in the experimental data, shown on Figure \ref{fig:lineouts}b, we can also see 
anisotropies in the flux distribution at 42 ns.

In the simulation, we can evaluate $\Delta v_\perp$ directly for each proton by measuring its spatial displacement on the recording screen
with respect to the point the proton would have landed with no plasma present, correcting for the spatial shift of the center of the pinhole
proton flux. This allows us to determine the RMS $\Delta v_\perp$ as experienced collectively by all protons passing through the pinhole.
We see in Figure \ref{fig:lineouts}c that FLASH predicts values of $\Delta v_\perp$ in very close agreement with the experimental ones. This
result further corroborates our previous explanation that the observed enhanced scattering is the result of a longer path-length through the
stochastic magnetic fields in the turbulent plasma, since the simulated proton beam is only subject to Lorentz forces and not other processes
(such as target charging or collisional broadening). In addition to the measurement of the RMS $\Delta v_\perp$, for the simulation we are able
to plot the probability distribution function (PDF) of deflection velocities; the result is shown in Figure \ref{fig:flash_pinhole}e.
As expected, we see a peak around the RMS value, in addition to a tail of higher values.

Finally, for the simulated data, we can apply the same analysis technique described and applied to the experimental data in the previous Section
to the synthetic pinhole-proton edge contours derived from FLASH. The results are shown in Figure \ref{fig:flash_pinhole}f. The
estimates obtained are comparable to the directly measured values. Due to the anisotropies present in the simulated pinhole-proton contours,
the uncertainty on the technique (calculated in the same way) is more significant than for the experimental data.

\section{Physical processes resulting in smearing of the proton beam}\label{a.smearing}

We find that most likely, the diffusive scattering is a result of stochastic magnetic fields.  Assuming classical velocity-space diffusion,
the estimate for the scattering velocity is obtained via a random-walk approximation, i.e., the sum of velocity deflections
$\delta v_{\ell_B}\approx \omega_g \ell_B$, where $\omega_g\equiv q_e B_{rms}/m_p$ is the gyrofrequency of the protons,
$B_{rms} \equiv \sqrt{\left<{\mathbf{B}^2}\right>}$ the measured RMS magnetic field, $q_e$ the electron charge, and $m_p$ the proton mass;
this results in $\Delta v_\perp \approx \omega_g \sqrt{\ell_i \ell_B}\approx q_e B_{rms} \sqrt{\ell_i \ell_B}/m_p$, where $\ell_i$ is the
size of the interaction region. As mentioned in the text, when we take $\ell_i \sim 0.8 \, \mathrm{mm}$, $B_{rms} \sim 100 \, \mathrm{kG}$
and $\ell_B \sim 50 \, \mu \mathrm{m}$, we find $\Delta v_\perp \sim 1.9 \times 10^7 \, {\rm cm}\,{\rm s}^{-1}$, which is indeed consistent
with the observed RMS deflection velocity in Figure \ref{fig:lineouts}c.

As mentioned in Section \ref{sec:experiments}, there are processes in addition to stochastic magnetic or electric fields which could result in the smearing
of features on the proton image. In this rest of this Section, we discuss estimates of these effects in more detail, and show that they are negligible. 

\subsection{Capsule size and proton velocity uncertainty}
The proton source, originating from an imploded capsule containing $\mathrm{D^3He}$ and $\mathrm{D}_2$ gas, has a finite source size, as well as
a spread in emitted proton energies.  The effective source size $d_c$ of the capsule for 15.0 MeV protons is 45 $\mu$m \citep{Li2006}. This is
smaller than the initial capsule size, because fusion reaction only happens in the hot region formed well inside the imploding capsule. Including
the magnification, $\mathcal{M}$, of the proton imaging diagnostic, we conclude that all structures with size below $\mathcal{M} d_c$ will be smeared
out. This is consistent with the slope at the edge of the pinhole image for the calibration shots seen in Figure \ref{fig:lineouts}a. The effect source size for 3.3 MeV
protons is believed to be somewhat larger, an observation born out in Figure \ref{fig:lineouts}b: the 3.3 MeV calibration profile is broader than the 15.0 MeV one. 

The blur resulting from uncertainties in the initial proton energy can be estimated as
\begin{equation}
\delta_v = r_{det} \delta \theta_B \frac{\delta V}{V} \approx 40 \textrm{ } \mu \textrm{m}. 
\label{eqn:veluncertainty}
\end{equation}
and as  $\delta_B \gg 40 \mu$m, this is negligible. The term $\delta V/V$ is calculated from the experimentally determined energy uncertainties
$\delta W/W$: these are $\sim$ 3\% and $\sim$ 4.5\% for D$^3$He and DD protons, respectively.  Using $\delta W/W \approx 2\delta V/V $, this gives
estimates for $\delta V/V$ as $\sim$ 1.5\% for 15.0 MeV protons and $\sim$ 2.25\% for 3.3 MeV.  Thus, this second effect is likely to be negligibly small. 

In short, given the relatively small size of diffusive scattering due to stochastic magnetic fields compared to the finite source size, accounting
for the source size is essential. As described in the previous Section, we carry out our analysis using experimental pinhole images created in the
absence of a plasma, and so automatically account for these effects. 
\begin{figure}[htbp]
%\begin{center}
\includegraphics[width=0.9\linewidth]{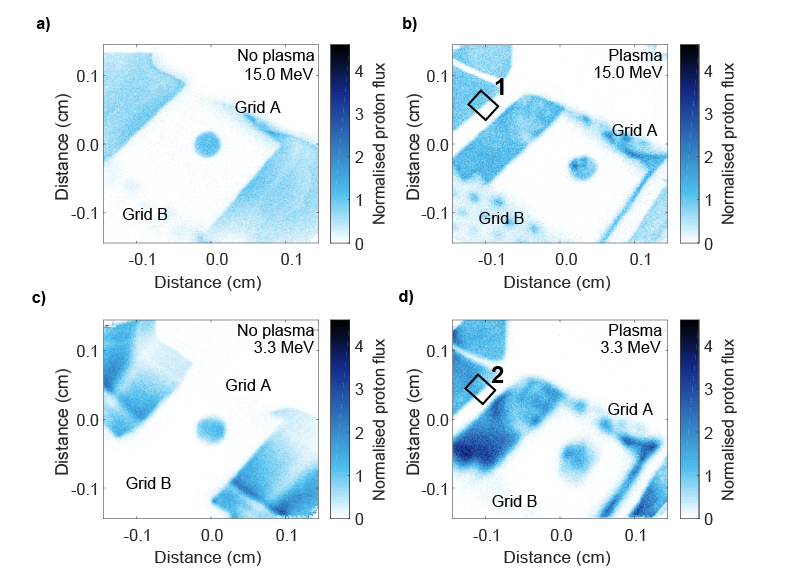}
\caption{\textbf{Charging of the target setup}.  Radially-averaged lineouts from the flux-weighted center of the pinhole in the case of no plasma present
in the interaction region are shown for \textbf{a)} 15.0 MeV and \textbf{c)} 3.3 MeV protons.  This is compared to lineouts taken across an edge section
of the target package from a proton radiograph in which there was a plasma present in the interaction region, demarcated by 1 and 2 for \textbf{b)}
15.0 MeV and \textbf{d)} 3.3 MeV protons.)}
\label{fig:prad_charging}
%\end{center}
\end{figure}

\begin{figure}[htbp]
\begin{tabular}{cc}
%\begin{center}
\includegraphics[width=0.45\linewidth]{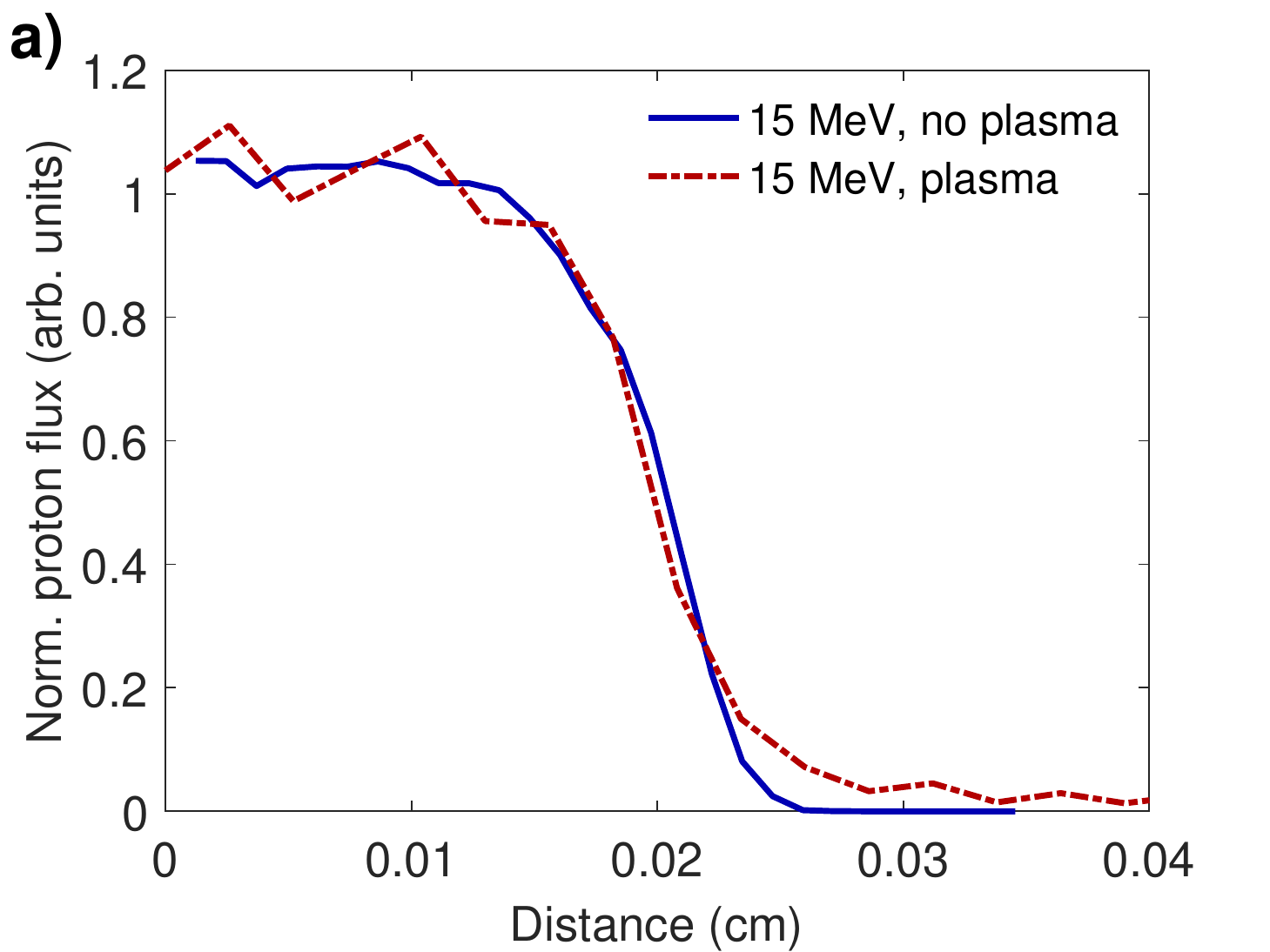} &
\includegraphics[width=0.45\linewidth]{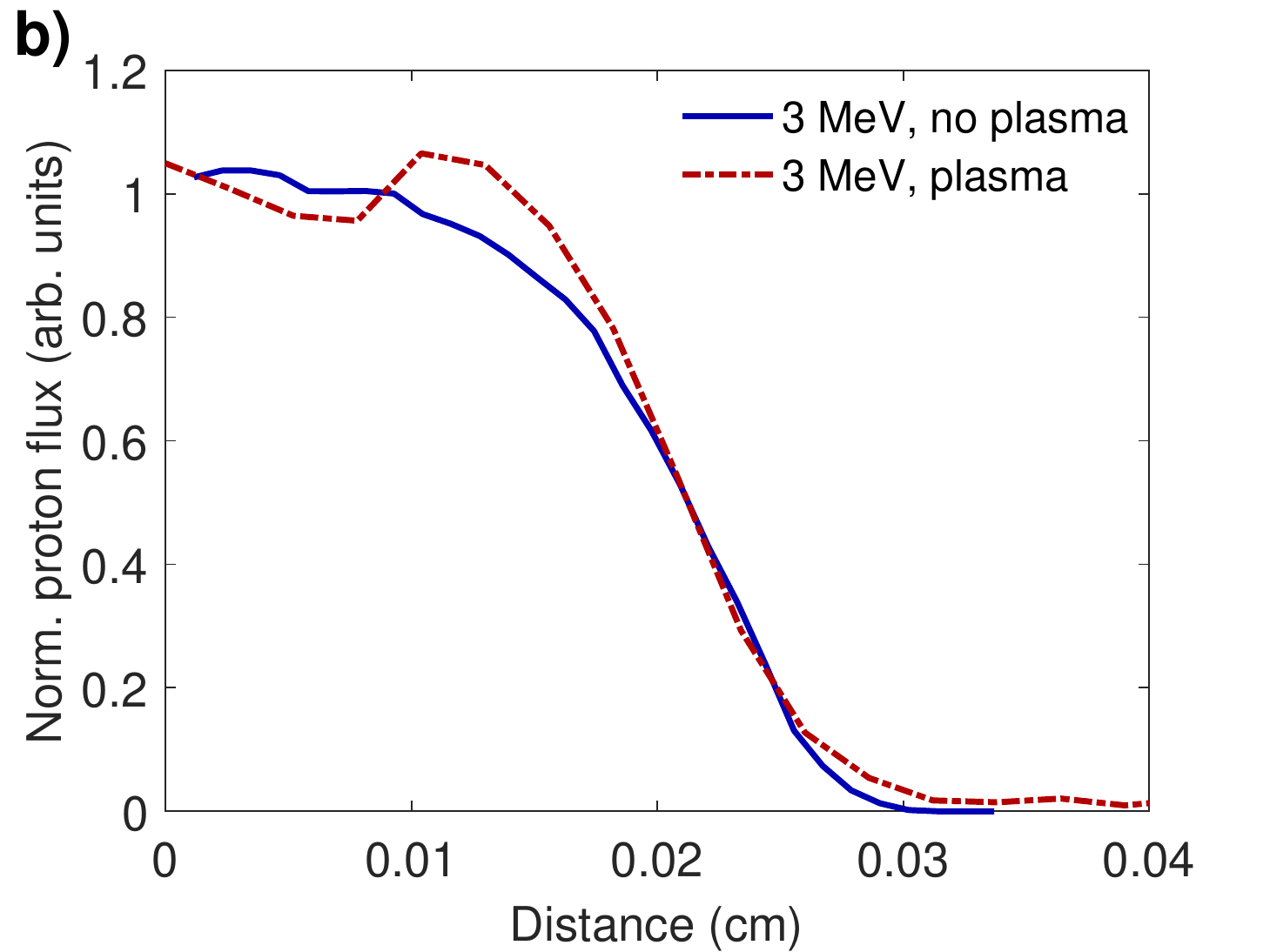}
\end{tabular}
\caption{\textbf{Charging effects on pinhole.} Lineouts across an edge section of the target package from a proton radiograph in which there was
a plasma present in the interaction region (red, dashed) vs a radially-averaged lineout of the pinhole in the case of no plasma present in the
interaction region (blue, solid) for \textbf{a)} 15.0 MeV and \textbf{b)} 3.3 MeV protons.  It is clear that the slopes of the two cases are
so similar that an electric field due to charging of the target is negligible.}
\label{fig:lineouts_charging}
%\end{center}
\end{figure}

\subsection{Electric fields due to charging of the target package}
In principle, it is possible that charging of the target package (due, for example to high energy electrons produced by laser-plasma processes)
could allow for the emergence of an electric field.  To rule out this effect we compare a flux lineout across an edge section of the target package
from a proton radiograph in which there was a plasma present in the interaction region (shown in Figure \ref{fig:prad_charging}b
and \ref{fig:prad_charging}d, labeled as 1 and 2 for each proton species) to a radially-averaged lineout from the flux-weighted center of the
pinhole in the case of no plasma present in the interaction region (shown in Figure \ref{fig:prad_charging}a and \ref{fig:prad_charging}c).
These lineouts, shown in Figure \ref{fig:lineouts_charging}, have comparable slopes, indicating any electric field due to charging is
negligible in our analysis.

\subsection{Electric fields generated by plasma fluid motions}

In Section \ref{sec:char}, we assert that scattering due to stochastic electric fields generated by turbulent motions of the electric field are small.
This estimate is made by combining \ref{eq:deltavE}  %that is 
% \begin{equation}
% \Delta v_E \approx \frac{q E_{rms} \sqrt{\ell_i \ell_E}}{m_p V} \, ,
% \end{equation}
with an appropriate estimate for $E_{rms} \sim u_{jet} B_{rms}$, and $\ell_E \sim \ell_B$. Then, it follows that
\begin{equation}
\Delta v_{E} \sim 1.6 \times 10^5 \left(\frac{u_{jet}}{2 \times 10^{7} \, \mathrm{cm}\,\mathrm{s}^{-1}}\right) \left(\frac{B_{rms}}{100 \, \mathrm{kG}}\right) \left(\frac{V}{2.5 \times 10^{9} \, \mathrm{cm}\,\mathrm{s}^{-1}}\right)^{-1} \left[\frac{\ell_i \ell_B}{4 \times 10^{-4} \, {\rm cm^2}}\right]^{1/2}\, \mathrm{cm}\,\mathrm{s}^{-1} \, ,
\end{equation}
as stated in Section \ref{sec:experiments}.

\subsection{Transverse diffusion via collisions}
As the plasma is neither cold nor in thermal equilibrium, it is not straightforward to do an analytical calculation of transverse
diffusion due to collisional scattering.  Using the classical approach of setting a background particle's Coulomb potential equal
to the kinetic energy of an incoming test particle produces estimates which can be off by a factor of the order of unity \citep{Trubnikov1965}.
A quantum mechanical treatment is required which also takes into account long-range interactions, and we therefore used SRIM \citep{Ziegler1999a}
to simulate the perpendicular deflection of 3.3 and 15 MeV protons through 3.7 mm (the largest reported $\ell_i$ at very late times) of a CH plasma
of density 9 $\times$ 10$^{19}$.  We found this deflection corresponds to $\Delta v_\perp = 2 \times 10^6\, \mathrm{cm}\,\mathrm{s}^{-1}$ for
DD protons and $\Delta v_\perp = 8 \times 10^5\, \mathrm{cm}\,\mathrm{s}^{-1}$ and D$^3$He protons.  Again, this is an order of magnitude lower
than the measured $\Delta v_\perp$, and is therefore negligible in our analysis.

\subsection{Electrostatic beam instabilities}

It is well known that, in general, a beam of protons can drive kinetic instabilities. To determine whether this could occur for our experiment,
we outline here an analysis investigating the kinetic stability of the combined plasma species and beam distribution functions.  Linear kinetic
stability for electrostatic modes can be established by considering the zeros of the dielectric function given by \citet{Boyd2003} and \citet{Krall1973}
\begin{equation}
\epsilon(p,\bm{k}) = 1 - \frac{\omega^2_{pe}}{k^2} \int_{C_L} dv_z \frac{d \bar{F}(v_z)/ dv_z}{v_z - i p/k} \, ,
\label{eqn:dielectric}
\end{equation}
where $p$ is complex frequency, $\omega_{pe}$ is the electron plasma frequency, $v_z$ is the velocity of a particle in the distribution in the
dimension the 1D distribution function, $C_L$ contour of interest in momentum space, the weighted distribution function, $\bar{F}$, is defined as
\begin{equation}
\bar{F} \equiv \frac{1}{n_e} \sum_{\alpha} Z^2_{\alpha} \frac{m_e}{m_{\alpha}} F_{\alpha}
\label{eqn:weightdist}
\end{equation}
for mass, $m$, atomic charge, $Z$, and distribution function, $F$ for all respective species, denoted by $\alpha$. 
It can then be shown that any isotropic, monotonically decreasing distribution function is inherently stable against infinitesimal (linear),
electrostatic perturbations.  

The plasma in our experiment is collisional and contains two ion species (carbon and hydrogen) in equal ratio. We therefore assume electron,
hydrogen, and carbon have the following respective Maxwellian distributions:
\begin{equation}
F_e(v_z) = \frac{n_e}{\pi^{1/2} v_{th,e}} \exp{\left[-\left(v_z/v_{th,e}\right)^2\right]} \, ,
\end{equation}
\begin{equation}
F_H(v_z) = \frac{n_{H}}{\pi^{1/2} v_{th,H}} \exp{\left[-\left(v_z/v_{th,H}\right)^2\right]} + \frac{n_b}{\pi^{1/2} \delta V} \exp \left[-((v_z-V)/\delta V)^2\right] \, ,
\end{equation}
\begin{equation}
F_C(v_z) = \frac{n_C}{\pi^{1/2} v_{th,C}} \exp{\left[-\left(v_z/v_{th,C}\right)^2\right]} \, ,
\end{equation}
where $v_{th,e} = \sqrt{2 T/m_e}$ is the thermal electron velocity, $v_{th,H} = \sqrt{2 T/m_p}$ the thermal plasma proton velocity,
$v_{th,C} = \sqrt{2 T/m_C}$ the plasma carbon ion velocity, $V$ the beam velocity, $\delta V$ the (thermal) width of the beam, $n_e$
the electron density, $n_H$ the plasma proton density, $n_b$ the beam proton density, and $n_C$ the plasma carbon ion density. We have
taken the ion and electron temperatures to be the same: $T_e = T_i = T$ \citep{Tzeferacos2017}.
Assuming quasineutrality, we have
\begin{equation}
n_e = n_H + 6 \, n_C \, .
\end{equation}
We have measured (see Appendix \ref{sec:a.plasma} on TS diagnostics) $n_e = 9 \times 10^{19} \, \mathrm{cm}^{-3}$, so $n_H = n_C = n_e/7 = 1.28 \times  10^{19} \, \mathrm{cm}^{-3}$.
The total proton count recorded on the CR-39 detector is 
$\sim 25,000$ for 15.0 MeV protons, and $\sim 50,000$ for 3.3 MeV protons. Since the proton pulse duration is $150 \, \mathrm{ps}$, this gives a particle
density in the proton beam of $n_b \approx 10^{8} \, \mathrm{cm}^{-3}$. 

We now take the weighted distribution using Equation \ref{eqn:weightdist} and examine $\frac{d\bar{F}}{dv_z}$ to see if a bump in $\bar{F}$ exists,
i.e., if $\frac{d\bar{F}}{dv_z}$ goes positive anywhere. If it does not, then the distribution function is monotonically decreasing. We have
\begin{eqnarray*}
\frac{d\bar{F}}{dv_z} &=& \frac{1}{n_e}\frac{n_e}{\pi^{1/2} v_{th,e}} \frac{-2v_z}{v_{th,e}^2} \exp{\left[-\left(v_z/v_{th,e}\right)^2\right]} + \frac{1}{n_e} \frac{m_e}{m_p} \frac{n_{H}}{\pi^{1/2} v_{th,H}}\frac{-2v_z}{v_{th,H}^2} \exp{\left[-\left(v_z/v_{th,H}\right)^2\right]} \\
& & + \frac{1}{n_e} \frac{m_e}{m_p}  \frac{n_b}{\pi^{1/2} \delta V} \frac{-2(v_z-V)}{\delta V^2} \exp \left[-((v_z-V)/\delta V)^2\right] \\ & & + \frac{1}{n_e} \frac{m_e}{m_{C}} \frac{n_{C}}{\pi^{1/2} v_{th,C}}\frac{-2v_z}{v_{th,C}^2} \exp{\left[-\left(v_z/v_{th,C}\right)^2\right]} \, .
\end{eqnarray*}
The distribution is clearly dominated by the electron distribution function.  Any bump in the weighted distribution function must occur near the
maximum of the beam distribution function -- that is, when $v_z \sim V$. For such $v_z$, the contributions to $\frac{d\bar{F}}{dv_z}$ from the
plasma proton and carbon species is exponentially small ($V \gg v_{th,H}, v_{th,C}$) and therefore negligible. Therefore, we can approximate
$\frac{d\bar{F}}{dv_z}$ for $v_z \sim V$ as
\begin{equation}
\frac{d\bar{F}}{dv_z} \approx \frac{-2v_z}{v_{th,e}^3} \exp{\left[-\left(v_z/v_{th,e}\right)^2\right]} +  \frac{n_b}{n_e} \frac{m_e}{m_p}\frac{2(V-v_z)}{\delta V^3} \exp \left[-((v_z-V)/\delta V)^2\right] \, .
\end{equation}
The global maximum of the second term on the right occurs when $v_z = V - v_{th,b}/\sqrt{2}$, where $v_{th,b}$ is the beam thermal
velocity (given by a spread in energies of about $\sim$ 280 eV), and can be recast as
\begin{equation}
\frac{n_b}{n_e} \frac{m_e}{m_p}\frac{2(V-v_z)}{\delta V^3} \exp \left[-((v_z-V)/v_{th,b})^2\right] = \frac{n_b}{n_e} \frac{m_e}{m_p} \frac{\sqrt{2}}{\delta V^2} \exp{\left(-1/2\right)} \, .
\end{equation}
Meanwhile, for $v_z \sim V$, the first term is 
$-2 V/v_{th,e}^3 \exp{\left[-\left(V/v_{th,e}\right)^2\right]}$. Taking the ratio of these terms, we find
\begin{equation}
\frac{d\bar{F}_b/{dv_z}}{|d\bar{F}_e/{dv_z}|} \sim \frac{n_b}{n_e} \frac{m_e}{m_p} \frac{v_{th,e}^2}{\delta V^2} \exp{\left[\left(V/v_{th,e}\right)^2\right]} \, ,
\end{equation}
where we emphasize that $d\bar{F}_e/{dv_z} < 0$. Substituting numerical values
$V = V_{\mathrm{DD}} \approx 2.5 \times 10^{9} \, \mathrm{cm}\,\mathrm{s}^{-1}$, $\delta V \approx 6.3 \times 10^{7} \, \mathrm{cm}\,\mathrm{s}^{-1}$, $v_{th,e} \approx 8.4 \times 10^{8} \, \mathrm{cm}\,\mathrm{s}^{-1}$, $n_b \approx 10^{8} \, \mathrm{cm}^{-3}$,
and $n_e \approx 10^{20} \, \mathrm{cm}^{-3}$, it is clear that
\begin{equation}
\frac{d\bar{F}_b/{dv_z}}{|d\bar{F}_e/{dv_z}|} \ll 1 \, ,
\end{equation}
a consequence of the twelve orders of magnitude difference between the plasma and beam densities. Clearly, the slope of the proton beam distribution
function is too small to overcome the negative slope of the electron distribution function.
Thus, we conclude that the beam density is too small to drive electrostatic kinetic beam instabilities. 

In conjunction with this conclusion, we do not expect scattering by unmagnetized plasma waves (i.e., via Landau damping) or any other electrostatic
effects to be the culprit for the rise in diffusion with time.  Firstly, the density and/or temperature variation from 32 ns to 42 ns is quite small
(i.e., the Debye length is roughly constant), whereas such an effect would require a significant increase in temperature and decrease in density to
promote additional scattering.  Secondly, the electric fields to support such wave amplification would be a significant fraction of the total energy
of the system, implying an electrostatic instability at play which, in fact, Landau damping would suppress in a plasma with near Maxwellian particle
species distributions such as those relevant for our experiment.

\section{Determination of the range of physical conditions for which diffusion is not affected by spatial intermittency}\label{sec:a.intermittency}

\begin{figure}[htp]
  \centering
  \includegraphics[width=0.95\linewidth]{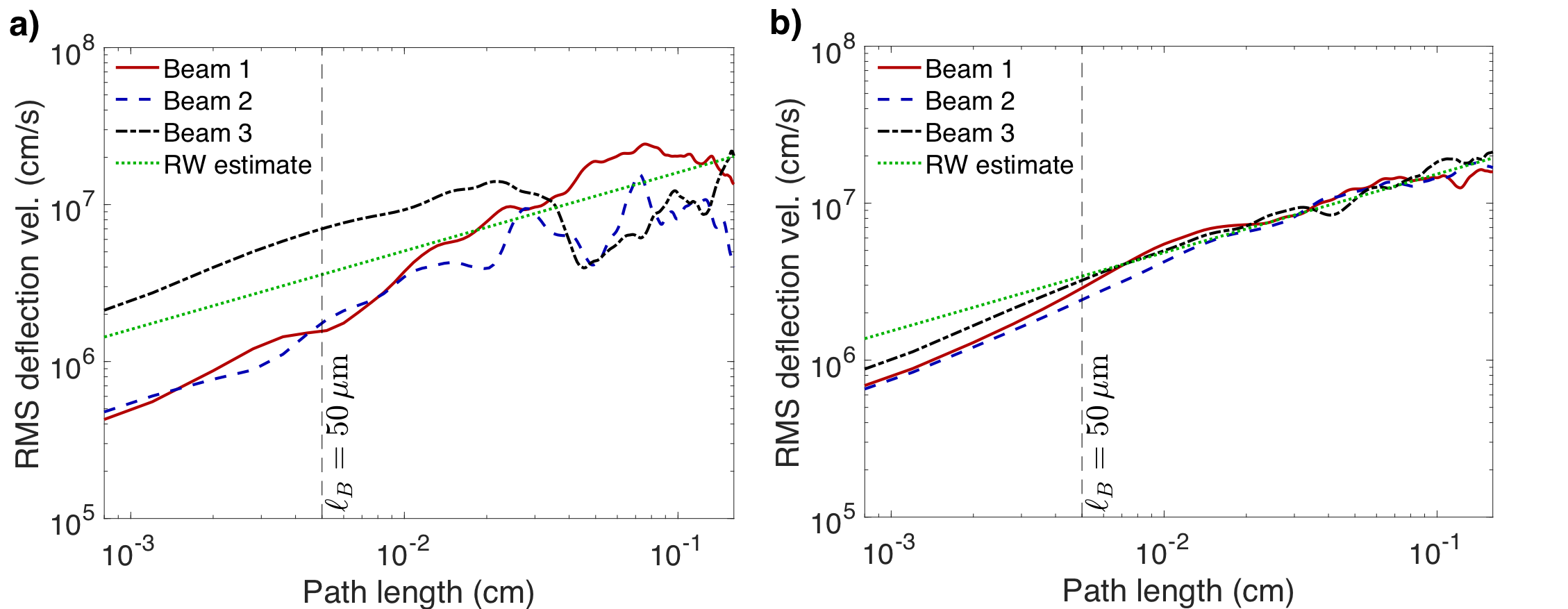}
  \caption{\textbf{Dependence of RMS deflection velocity on path length and beam cross-sectional area}. \textbf{a)} RMS
  deflection velocity $\Delta v_{\perp}$ acquired by beams of 15.0 MeV protons (total number $5\times 10^5$) with cross-sectional
  area $A = 1 \times 10^{-5} \, \mathrm{cm}^2$ passing through a prescribed stochastic magnetic field, as a function of path length.
  The full stochastic magnetic field -- which has non-Gaussian, exponential statistics, as well as $B_{rms} = 80 \, \mathrm{kG}$,
  and $\ell_B = 50 \, \mu \mathrm{m}$ -- is generated using the approach described in~\cite{Bott2017} on a $201^3$ three-dimensional array,
  with (physical) side length $\ell_i = 0.16 \, \mathrm{cm}$. To calculate $\Delta v_{\perp}$ at a given path length, the proton beam
  is propagated using the small-deflection approximation from one side through the whole array, with the RMS deflection velocity recorded
  as the particle propagates. The initial velocity is set to be exactly perpendicular to the side on which the beam is incident. Three
  different beams are employed (`Beam 1, 2 and 3'), with randomly chosen centroids. The random-walk (`RW') estimate is calculated using
  the known properties of the field and Equation \ref{eq:deltavB}. \textbf{b)} Same as a), except the beams are given cross-sectional
  areas $A = 9 \times 10^{-4} \, \mathrm{cm}^2$. For $\ell_i > \ell_B$ we see the expected convergence to Equation \ref{eq:deltavB}.}
\label{fig:beam_area_test}
\end{figure}

As stated in Section \ref{sec:discussion}, the FLASH simulations and our experimental results demonstrate that the diffusion of charged particles in
the large-$r_g/\ell_B$ regime is not affected by the spatial intermittency of the stochastic magnetic fields for the physical conditions
we studied.  Here we describe numerical simulations we did to determine the range of physical conditions for which diffusion is not affected
by spatial intermittency.  In the simulations, test particle beams propagate through a given stochastic magnetic field, measuring the RMS
deflection velocity as a function of path length.

For simplicity's sake, we do not use the FLASH-simulated fields, but instead consider homogeneous, single-scale stochastic fields with a
prescribed magnetic-energy spectrum, chosen so that the correlation length $\ell_B$ and RMS magnetic-field strength $B_{rms}$ are the same
as those found in our experiment. The results show that for a beam with a cross-sectional area $A$ much smaller than the characteristic
cross-sectional areas of magnetic structures (i.e., $A/4 \ell_B^2 \ll 1$) -- which is effectively analogous in our parameter regime to
that of a single particle -- the three beams roughly obey Equation \ref{eq:deltavB} for path lengths $\ell_i$ greater than the magnetic
field coherence length $\ell_B$, but there are significant deviations from the exact result as a function of path length and beams with
distinct initial locations have different behaviors. See, e.g., Figure~\ref{fig:beam_area_test}a.  In contrast, beams with
cross-sectional areas satisfying $A/4 \ell_B^2 > 1$ converge to Equation \ref{eq:deltavB} for $\ell_i > \ell_B$, as expected.  See, e.g.,
Figure~\ref{fig:beam_area_test}b.  These results show that the use of standard diffusion theory to model the transport of
charged particles applies when the magnetic field is weak, turbulent, and spatially intermittent; $r_g \gg \ell_B$; and the cross-sectional
area of the beam $A$ and the path length $\ell_i$ are greater than the magnetic field coherence length $\ell_B$.  The transport of ultra-high-energy
cosmic rays in the intergalactic medium satisfies these conditions, so our experiment validates the use of standard diffusion theory to model it,
e.g., \citet{Kotera2008,Globus2008,Globus2017,Globus2019}.

\bibliography{LauraOxfordBib2a,citationsPetros,LauraOxfordBib2si,Petros}{}
\bibliographystyle{aasjournal}

\end{document}